\documentclass[10pt]{article}
\usepackage{a4wide}
\usepackage{setspace}
\usepackage{pdflscape}
\usepackage{amsmath} 
\usepackage{amssymb} 
\usepackage{subfigure} 
\usepackage{graphicx}
\usepackage{framed}
\usepackage{natbib}
\graphicspath{{figures/}}
\usepackage{float}
\usepackage[small]{caption}
\usepackage{rotating}
\usepackage{booktabs}
\usepackage{url}
\usepackage{enumerate}
\usepackage{authblk}
\usepackage{graphicx}

\floatstyle{ruled}
\newfloat{algorithm}{htbp}{lop}
\floatname{algorithm}{Box}

\RequirePackage[OT1]{fontenc}
\RequirePackage{amsthm,amsmath,graphicx}
\RequirePackage[colorlinks,citecolor=blue,urlcolor=blue]{hyperref}
\RequirePackage{bm}

\def\tr{\text{tr}}
\def\det{\text{det}}

\begin{document}


\title{Distance-based analysis of variance: approximate inference and an application to genome-wide association studies}

\author{Christopher Minas and Giovanni Montana\footnote{Corresponding author: {\tt g.montana@imperial.ac.uk}} \footnote{Data used in preparation of this article were obtained from the Alzheimer’s Disease Neuroimaging Initiative (ADNI) database (adni.loni.ucla.edu). As such, the investigators within the ADNI contributed to the design and implementation of ADNI and/or provided data but did not participate in analysis or writing of this report. A complete listing of ADNI investigators can be found at : \url{http://adni.loni.ucla.edu/wp-content/uploads/how_to_apply/ADNI_Acknowledgement_List.pdf}} \\Department of Mathematics\\Imperial College London}

\maketitle

\begin{abstract}
In several modern applications, ranging from genetics to genomics and neuroimaging, there is a need to compare observations across different populations, such as groups of healthy and diseased individuals. The interest is in detecting a group effect. When the observations are vectorial, real-valued and follow a multivariate Normal distribution, multivariate analysis of variance (MANOVA) tests are routinely applied. However, such traditional procedures are not suitable when dealing with more complex data structures such as functional (e.g. curves) or graph-structured (e.g. trees and networks) objects, where the required distributional assumptions may be violated. In this paper we discuss a distance-based MANOVA-like approach, the DBF test, for detecting differences between groups for a wider range of data types. The test statistic, analogously to other distance-based statistics, only relies on a suitably chosen distance measure that captures the pairwise dissimilarity among all available samples. An approximate null probability distribution of the DBF statistic is proposed thus allowing inferences to be drawn without the need for costly permutation procedures. Through extensive simulations we provide evidence that the proposed methodology works well for a range of data types and distances, and generalizes the traditional MANOVA tests. We also report on the  analysis of a multi-locus genome-wide association study of Alzheimer's disease, which has been carried out using several genetic distance measures. The DBF test has detected causative genetic variants previously known in the literature.
\end{abstract}

\section{Introduction}


        


Multivariate analysis of variance (MANOVA) techniques are routinely applied to compare real-valued multivariate observations collected in two or more populations and test whether the mean vectors in those groups have been drawn from the same underlying sampling distribution (see \cite{anderson1984introduction} and \cite{krzanowski2000pma}, for example). These methods are applicable in a wide range of scientific areas. For instance, in genomics, microarrays allow the researcher to investigate the behaviour of thousands of genes simultaneously under various conditions. A common task consists of observing the expression levels of a group of genes sharing the same biological function, and assessing whether the mean expression levels in these gene sets differ across experimental conditions. Compared to a univariate approach involving only one gene expression measurement at a time, a MANOVA test may capture potential gene interactions and hence provide a more powerful and biologically meaningful way of detecting subsets of differentially expressed genes \citep{Szabo2003, Tsai2009, Xu2008, Shen2011}.

In several applications, however, MANOVA tests may be inappropriate for at least two main reasons. Firstly, when the observations are multivariate, the multivariate normality assumption may not necessarily hold, e.g., if the observations are heavily skewed or are discrete-valued. 
Secondly, the observations may not necessarily be represented by vectorial data structures. In an increasingly large number of applications, the data being compared across groups are functional (e.g. curves) and graph-structured (e.g. trees and networks) objects. This is indeed the case with several applications we have a particular interest in, of which the following are three representative examples:

\begin{itemize}
\item In genome-wide association (GWA) studies, the allele frequencies observed at multiple genetic markers (single nucleotide polymorphisms, or SNPs) in healthy and diseased subjects are compared in order to identify genetic variants that are associated with disease risk. Although GWAs have emerged as popular methodologies for gene mapping, traditional analyses that rely on testing one individual marker at a time are believed to have low power to detect small genetic effects. They have also been shown to have limited reproducibility, and are unable to detect effects which are due to gene-gene interactions. For these reasons, sets of multiple markers (i.e. markers observed at multiple loci) can be used instead, giving rise to discrete-valued vectors to be compared across groups \citep{wu2010powerful}.  
\item In longitudinal microarray experiments, repeated measurements of a gene's expression level are recorded at different time-points, and the resulting time course is often modelled as a functional object or curve, i.e. an infinite-dimensional data structure \citep{Storey2005} (see Figure \ref{m_tub_gene_curves} for an example of such a functional dataset). The resulting gene curves observed under two or more experimental conditions are then compared using a test statistic that accounts for the functional nature of the data \citep{Berk2012, Berk2009a}. 
\item In neuroimaging studies, there is increasing interest in comparing individual brain connectivity networks inferred in subjects that belong to two or more clinically relevant groups, e.g. diseased and healthy individuals. A network is a collection of nodes (vertices) and links (edges) between pairs of nodes. Such brain connectivity networks describe the set of connections in the neural system, or {\it connectome}, in which nodes could be neurons or cortical areas, and edges could be axons or fibre tracts. Thus, edges could refer to the {\it structural} connectivity of a neural network, or could signify correlations between the activity patterns of nodes forming {\it functional} connectivity \citep{Rubinov2010}. Although univariate test statistics can be used to compare particular summary features extracted from the graphs, such as centrality measures, there is a need to compare individual graphs directly.     
\end{itemize}

Since standard MANOVA techniques assume real-valued vectorial data representations to compare mean vectors, they are not always suitable. Moreover, in many of these applications, a very large number of tests are needed to be simultaneously performed within the same experiment. For instance, in genomic experiments such as those involving gene sets or time courses, as well as GWA studies, the number of comparisons can range from many tens of thousands to millions. Computationally efficient inferential methods are therefore required.


In this article we consider the problem of comparing two or more independent random samples, representative of some underlying populations, by using a distance-based analysis of variance approach. As in traditional MANOVA, the total variability observed in the random samples is decomposed into the sum of between-group and within-group variability. The decomposition only relies on a suitable distance measure defining how dissimilar any two samples are, and does not require the computation of means in each group; both attractive properties when dealing with non-vectorially structured objects. For a given application, a suitable distance measure is defined depending on the nature of the data (whether vectorial or not, for example), and on the specific objectives of the study, since each distance measure captures a different feature of the data. We discuss particular distances for genetic and functional data later in this article. Pairwise distances between all available samples are computed and stored in a square, symmetric distance matrix, with samples deemed more similar if the distances between them are small.  


A number of distance-based tests for no group effect have been proposed in the literature as generalizations of classical MANOVA procedures, and are applicable when the MANOVA assumptions are violated, such as with data of the type described above. These include the Mantel test \citep{mantel1967detection}, the MRPP test \citep{mielke2007permutation}, and the more recent DBF test \citep{minas2011distance}. Each statistic utilizes some notion of within- and between-group variability in terms of distances: Mantel considers between-group variability, MRPP considers within-group variability, and DBF considers the ratio of between- to within-group variability. 
An important, and in many cases limiting, feature of such distance-based tests is that exact distribution theory of the statistics under the null hypothesis of no group effect is unavailable. This is due to the wide variety of distance measures available for different data types leading to different distributional characteristics of the distances \citep{mantel1967detection}. Due to this limitation, non-parametric and computationally intensive statistical procedures based on permutations are typically applied, since they do not rely on any distributional assumptions. The only requirement is that samples are exchangeable across groups under the null hypothesis of no group effect \citep{pesarin2010permutation}. Statistical significance of an observed statistic is assessed by comparing the estimate against the discrete sampling distribution obtained by permuting the samples across groups and recomputing the statistic each time. For each permutation, this is equivalent to permuting the rows and columns of the corresponding distance matrix simultaneously, and recomputing the statistic with the permuted distance matrix \citep{mielke2007permutation,minas2011distance}. Thus the distance matrix needs only be computed once, and inferences can be drawn based on this matrix.

Despite being routinely adopted, such a permutation approach suffers from some important limitations. Even for low sample sizes, exact p-values require a very large number of permutations to be obtained; for instance, for $12$ observations, the total number of permutations required is $O(10^8)$. Exact p-values are often impossible to obtain due to computational and time constraints, and Monte Carlo procedures are used instead whereby a smaller number of randomly chosen permutations are used to approximate the p-values. This approach inevitably introduces sampling errors \citep{berry1983moment}. Moreover, whereas large p-values can be well approximated by a Monte Carlo approach, smaller ones will be estimated less accurately \citep{mielke2007permutation,knijnenburg2009fewer}. For example, in order to obtain a permutation p-value within $10^{-5}$ of the true p-value, it has been shown that $O(10^7)$ permutations are required. In several real applications, we have observed that less than $O(10^5)$ permutations are typically used, especially when a very large number of statistical tests must be performed simultaneously. In these cases, the total number of permutations per test may be vastly limited even when ad-hoc parallel implementations that run on high-performance computing facilities are being used. 


In order to deal with these limitations, in this work we propose a closed-form, continuous approximation to the null sampling distribution of the DBF statistic \citep{minas2011distance}, which was originally proposed for the analysis of functional data arising from longitudinal microarray experiments. Our main contribution here is to enable distance-based analysis of variance testing in a wide range of applications, in a computationally efficient manner without the need for permutations. Using extensive simulations, we demonstrate that the proposed methodology works well for several different data structures, including functional and discrete-valued vectorial data,  in addition to several distances, and that it provides a good approximation of the permutation p-values. We also show that traditional ANOVA and MANOVA tests are special cases of the proposed approach when their assumptions are satisfied. For these special cases we show that the proposed continuous distribution approximates the true distributions well. The computational cost of analyzing large datasets is thus reduced significantly without compromising the accuracy of the results. 

A second contribution of this work consists of applying of the proposed methodology to a multi-locus GWA study of the Alzheimer's Disease Neuroimaging Initiative (ADNI) cohort (see for instance, \cite{vounou2010discovering, Vounou2012}). To the best of our knowledge, this is the first multi-locus case-control genetic study on this cohort. By using a range of genetic distances we show that the DBF test identifies previously known genetic variants associated with Alzheimer's disease, and discuss a comparison with a distance-based method that has been specifically designed for the analysis of such large-scale case-control studies. 

This article is organized as follows. In Section \ref{dbtest}, we briefly review the classical variance decomposition and MANOVA tests, followed by a detailed account of the proposed distance-based variance decomposition. The corresponding DBF statistic is defined, and theoretical connections with classical MANOVA and ANOVA tests are included. Section \ref{db_bg_var} describes how the sampling distribution of the DBF statistic exhibits skewed characteristics, and we justify the use of the Pearson type III distribution for its approximation. Approximate distribution theory for the DBF statistic then follows in Section \ref{null_dbf_dist}. We provide evidence that the approximation works well for a range of distances and data types in Section \ref{sims}, including empirical comparisons with ANOVA and MANOVA. Finally, in Section \ref{adni_gwas} we showcase the generality and applicability of the proposed approach by analyzing the ADNI dataset and reporting on the genetic variants we have identified. Conclusive remarks are given in Section \ref{conclusion}.




\section{A distance-based test of equality between populations}\label{dbtest}

\subsection{Traditional variance decomposition and MANOVA tests}\label{classical_var_decomps}

We first consider the case of $N$ independent vectorial and real-valued observations $\{\bm{y}_i\}_{i=1}^N$ in $\mathbb{R}^P$ that belong to one of $G$ populations with means $\{\bm{\mu}_g\}_{g=1}^G$. Each group is of size $N_g$ such that $N=\sum_{g=1}^G N_g$. We are interested in testing the null hypothesis that all the population means are equal, versus the alternative hypothesis that some of the means are not equal. 

Several standard statistical tests are based on the multivariate analysis of variance (MANOVA). The overall sample mean is given by $\bar{\bm{y}}=\frac{1}{N}\sum_{i=1}^N\bm{y}_i$ and each within-group sample mean is given by $\bar{\bm{y}}_{g}=\frac{1}{N_g}\sum_{i=1}^N\bm{y}_iI_{gi}$ for $g=1,\ldots,G$ where $I_{gi}$ is an indicator variable taking the value $1$ if observation $\bm{y}_i$ is in group $g$ and $0$ otherwise. The $P\times P$ total sum of squares matrix is given by 
\begin{equation}
\bm{T}=\sum_{i=1}^N\left(\bm{y}_i-\bar{\bm{y}}\right)\left(\bm{y}_i-\bar{\bm{y}}\right)^T,\nonumber
\end{equation}
and can be partitioned into the sum of between- and within-group sum of squares matrices,  
$$
\bm{B}=\sum_{g=1}^GN_g\left(\bar{\bm{y}}_{g}-\bar{\bm{y}}\right)\left(\bar{\bm{y}}_{g}-\bar{\bm{y}}\right)^T
\quad \text{ and } \quad
\bm{W}=\sum_{g=1}^G \sum_{i=1}^{N}\left(\bm{y}_{i}-\bar{\bm{y}}_{g}\right)\left(\bm{y}_{i}-\bar{\bm{y}}_{g}\right)^TI_{gi},
$$ 
respectively.


Existing MANOVA test statistics make use of different elements of this total sum of squares decomposition. For $G=2$, the well-known Hotelling's $T^2$ statistic is given by $N_1N_2(N-2)(\bar{\bm{y}}_1-\bar{\bm{y}}_2)^T\bm{W}^{-1}(\bar{\bm{y}}_1-\bar{\bm{y}}_2)/N$. For $G>2$, MANOVA statistics include Wilks' Lamda, $\det(\bm{W})/\det(\bm{T})$, the Pillai trace, $\tr\left(\bm{T}^{-1}\bm{B}\right)$, and the Lawley-Hotelling trace, $\tr\left(\bm{W}^{-1}\bm{B}\right)$ \citep{rencher2002methods,krzanowski2000pma}. 
Under the assumption that the observations are independent and identically distributed from a multivariate Normal distribution with mean $\bm{\mu}$ and variance-covariance matrix $\bm{\Sigma}$, some distributional results are available. For instance, for $G=2$, Hotelling's $T^2$ statistic (multiplied by a constant depending on $N$ and $P$) has an exact F distribution under the null;
\begin{equation}
\frac{N-P-1}{(N-2)P}T^2\sim  F_{P,N-P-1},\nonumber
\end{equation}
where $F_{P,N-P-1}$ denotes the F distribution with degrees of freedom $P$ and $N-P-1$. 
For $G>2$, Wilks' Lamda, the Pillai trace and the Lawley-Hotelling trace can all be similarly transformed to statistics which are well-approximated by the F distribution with degrees of freedom dependent on $N$, $G$ and $P$ (see, for example, \cite{rencher2002methods}). 

When $N<P$, as is typically the case with genomic datasets, the classical MANOVA tests cannot be applied directly. This is because the $\bm{T}$ and $\bm{W}$ matrices are singular, and so have at least one zero-valued eigenvalue and cannot be inverted. 
Several high-dimensional MANOVA settings have been considered in the literature, and tests of equality between groups have been proposed, some using traditional MANOVA statistics with generalized inverses (see \cite{srivastava2007multivariate} and \cite{schott2007some} for good reviews, and \cite{Tsai2009} for an application to gene expression data). One of the first tests was proposed by Dempster for $G=2$ \citep{dempster1960significance}, where an F-type statistic defined as the ratio of within- to between-group variability was proposed. 
This statistic was generalized for $G>2$ and named the Dempster trace criterion four decades later by \cite{fujikoshi2004asymptotic}. They noticed that the trace operator could be applied to the $\bm{B}$ and $\bm{W}$ sum of squares matrices to yield equivalent expressions to those proposed by Dempster. Although not stated explicitly, they summarize the total variability exhibited by the $P$ variables by $\tr(\bm{T})$. They then partition this into between- and within-group components via $\tr(\bm{T})=\tr(\bm{B})+\tr(\bm{W})$, which follows by applying the trace operator to the decomposition $\bm{T}=\bm{B}+\bm{W}$. The Dempster trace criterion was then defined as $\tr(\bm{B})/\tr(\bm{W})$, and a transformation of this statistic was shown to be asymptotically Gaussian. In the next section we describe a further elaboration of this approach which only relies on pairwise distances among samples.

\subsection{The distance-based variance decomposition}\label{dbv} 

As before, we consider $N$ independent observations $\{y_i\}_{i=1}^N$ belonging to $G$ groups each of size $N_g$ for $g=1,\ldots,G$. Now, however, we place no restriction on the nature of these observations; they can be of any form, for example, scalar-valued, vector-valued, functional, graph-structured, or images, amongst others. The fundamental assumption is that we are able to define a distance measure $d$, which may be either semimetric or metric, which quantifies the dissimilarity between any pair of observations in the random sample. All pairwise distances are then arranged in an $N\times N$ distance matrix $\bm{\Delta}=\left\{d(y_i,y_j)\right\}_{i,j=1}^N$. The choice of distance depends on the type of data and scientific problem at hand. We are then interested in testing the null hypothesis that there is no difference between the observations from different groups with respect to the chosen distance measure $d$, against the alternative that a difference exists between any two groups. That is, under the null, all observations are assumed to belong to the same population. 

First, we introduce a generalization of the variance decomposition approach using pairwise distances. Consider the quantity $\tr(\bm{T})$ associated with a set of vectorial observations. It is a measure of spread found by summing the squared Euclidean distance of each observation to the population mean vector. This quantity can be equivalently written using only pairwise Euclidean distances between observations, as follows:
\begin{eqnarray}
\tr(\bm{T})&=&\sum_{i=1}^N(\bm{y}_i-\bar{\bm{y}})^T(\bm{y}_i-\bar{\bm{y}}),\nonumber\\
&=&\frac{1}{2}\sum_{i=1}^N\bm{y}^T_i\bm{y}_i+\frac{1}{2}\sum_{j=1}^N\bm{y}^T_j\bm{y}_j-\frac{1}{N}\sum_{i=1}^N\sum_{i=1}^N\bm{y}_i^T\bm{y}_j\nonumber\\
&=&\frac{1}{2N}\sum_{i=1}^N\sum_{j=1}^N(\bm{y}_i-\bm{y}_j)^T(\bm{y}_i-\bm{y}_j)\nonumber\\
&=&\frac{1}{2N}\sum_{i=1}^N\sum_{j=1}^N d^2_E(\bm{y}_i,\bm{y}_j),\nonumber
\end{eqnarray}
where $d_E$ denotes the Euclidean distance. Thus, $\tr(\bm{T})$ is proportional to the sum of squared inter-point Euclidean distances between all $N$ observations. This well-known connection shows that the total variability of a given set of vectorial observations, traditionally found using the population mean, can be computed using only the inter-point Euclidean distances  \citep{gower1999analysis,anderson2001new,minas2011distance}. In an analogous manner, the within- and between-group variability quantities $\tr(\bm{W})$ and $\tr(\bm{B})$ can also be written in terms of squared Euclidean distances, as follows:
\begin{eqnarray}
\tr(\bm{W}) &=& \sum_{g=1}^G \sum_{i=1}^{N}\left(\bm{y}_{i}-\bar{\bm{y}}_{g}\right)^T\left(\bm{y}_{i}-\bar{\bm{y}}_{g}\right)I_{gi}\nonumber\\
&=&\sum_{g=1}^G\left(\frac{1}{2}\sum_{i=1}^N\bm{y}^T_i\bm{y}_iI_{gi}+\frac{1}{2}\sum_{j=1}^N\bm{y}^T_j\bm{y}_jI_{gj}-\frac{1}{N_g}\sum_{i=1}^N\sum_{i=1}^N\bm{y}_i^T\bm{y}_j I_{gi}I_{gj}\right)\nonumber\\
&=&\sum_{g=1}^G\left(\frac{1}{2N_g}\sum_{i=1}^N\sum_{j=1}^N\left(\bm{y}_i-\bm{y}_j\right)^T\left(\bm{y}_i-\bm{y}_j\right)I_{gi}I_{gj}\right)\nonumber\\
&=&\frac{1}{2}\sum_{g=1}^G\sum_{i=1}^N\sum_{j=1}^Nd^2_E(\bm{y}_i,\bm{y}_j)\frac{I_{gi}I_{gj}}{N_g},\nonumber
\end{eqnarray}
and since $\tr(\bm{B})=\tr(\bm{T})-\tr(\bm{W})$, we obtain
\begin{eqnarray}
\tr(\bm{B})&=&\frac{1}{2N}\sum_{i=1}^N\sum_{j=1}^N d^2_E(\bm{y}_i,\bm{y}_j) - \frac{1}{2}\sum_{g=1}^G\sum_{i=1}^N\sum_{j=1}^Nd^2_E(\bm{y}_i,\bm{y}_j)\frac{I_{gi}I_{gj}}{N_g}\nonumber\\
&=&\frac{1}{2N}\sum_{i=1}^N\sum_{j=1}^Nd^2_E(\bm{y}_i,\bm{y}_j)\left(1-\sum_{g=1}^G N\frac{I_{gi}I_{gj}}{N_g}\right).\nonumber
\end{eqnarray}

Generalizations of these quantities can be defined by replacing the Euclidean distance, $d_E$, with any distance $d$. Thus we can define the total variability of a set of observations $\{y_i\}_{i=1}^N$ with respect to distance $d$ as
\begin{equation}
T_{\bm{\Delta}}=\frac{1}{2N}\sum_{i=1}^N\sum_{j=1}^N d^2(y_i,y_j),\nonumber
\end{equation}
the within-group variability with respect to $d$ as
\begin{equation}
W_{\bm{\Delta}}=\frac{1}{2}\sum_{g=1}^G\sum_{i=1}^N\sum_{j=1}^N d^2(y_i,y_j)\frac{I_{gi}I_{gj}}{N_g},\nonumber
\end{equation}
and the between-group variability with respect to $d$ as 
\begin{equation}
B_{\bm{\Delta}}=\frac{1}{2N}\sum_{i=1}^N\sum_{j=1}^N d^2(y_i,y_j)\left(1-\sum_{g=1}^G N\frac{I_{gi}I_{gj}}{N_g}\right).\nonumber
\end{equation}

\noindent The total variability in the data captured by $T_{\bm{\Delta}}$ can hence  be decomposed into the sum of two components quantifying within- and between-group variability. That is, $T_{\bm{\Delta}}=W_{\bm{\Delta}}+B_{\bm{\Delta}}$, analogously to the decomposition $\tr(\bm{T})=\tr(\bm{W}) + \tr(\bm{B})$.

We can write $T_{\bm{\Delta}}$, $B_{\bm{\Delta}}$ and $W_{\bm{\Delta}}$ more compactly in matrix form by introducing the $N\times N$ Gower's centered inner product matrix \citep{gower1966sdp}, defined as
\begin{equation} 
\label{gmatrix}
\bm{G}_{\bm{\Delta}}=\left(\bm{I}_N-\frac{1}{N}\bm{J}_N\right)\left(-\frac{1}{2}\bm{\Delta}^2\right)\left(\bm{I}_N-\frac{1}{N}\bm{J}_N\right),\nonumber
\end{equation}
where $\bm{I}_N$ is the identity matrix of size $N$, $\bm{J}_{N}$ is the square matrix of ones of size $N$ and $\left(\bm{I}_N-\frac{1}{N}\bm{J}_N\right)$ is the centering matrix. This matrix contains all the information on the inter-point distances between the $N$ observations, and is such that its trace equals $T_{\bm{\Delta}}$; 
\begin{eqnarray}
\tr\left(\bm{G}_{\bm{\Delta}}\right)&=&\tr\left(\left(\bm{I}_N-\frac{1}{N}\bm{J}_N\right)\left(-\frac{1}{2}\bm{\Delta}^2\right)\right)\nonumber\\
&=&\tr\left(-\frac{1}{2N}\bm{\Delta}^2+\frac{1}{2N}\bm{J}_N\bm{\Delta}^2\right)\nonumber\\
&=&\frac{1}{2N}\tr\left(\bm{J}_N\bm{\Delta}^2\right)\nonumber\\
&=&\frac{1}{2N}\sum_{i=1}^N\sum_{j=1}^N d^2(y_i,y_j).\nonumber
\end{eqnarray}
Therefore we rewrite $T_{\bm{\Delta}}$ more conveniently as $\tr(\bm{G}_{\bm{\Delta}})$. For $W_{\bm{\Delta}}$ and $B_{\bm{\Delta}}$ we define the centered $N\times N$ matrix of constants encoding group membership of each observation to one of the $G$ groups as  
\begin{equation}
\bm{H}_c=
\begin{pmatrix}\frac{1}{N_1}\bm{J}_{N_1}&&&0\\
&\frac{1}{N_2}\bm{J}_{N_2}&&\\
&&\ddots&\\
0&&&\frac{1}{N_G}\bm{J}_{N_G}\end{pmatrix}-\frac{1}{N}\bm{J}_N,\nonumber
\end{equation}
where $\bm{J}_{a}$ is the square matrix of ones of size $a$. Since this matrix is centered, we have that  
\begin{equation}
\left(\bm{I}_N-\frac{1}{N}\bm{J}_N\right)\bm{H}_c\left(\bm{I}_N-\frac{1}{N}\bm{J}_N\right)=\bm{H}_c,\nonumber
\end{equation}
and we use this fact in the evaluation of the quantity $\tr(\bm{H}_c\bm{G}_{\bm{\Delta}})$ to derive expressions for $W_{\bm{\Delta}}$ and $B_{\bm{\Delta}}$ in terms of $\bm{G}_{\bm{\Delta}}$. We have
\begin{eqnarray} 
\tr(\bm{H}_c\bm{G}_{\bm{\Delta}})&=&\tr\left(\left(\begin{pmatrix}\frac{1}{N_1}\bm{J}_{N_1}&&&0\\
&\frac{1}{N_2}\bm{J}_{N_2}&&\\
&&\ddots&\\
0&&&\frac{1}{N_G}\bm{J}_{N_G}\end{pmatrix}-\frac{1}{N}\bm{J}_N\right)\left(-\frac{1}{2}\bm{\Delta}^2\right)\right)\nonumber\\
&=&\frac{1}{2N}\tr\left(\bm{J}_N\bm{\Delta}^2\right)-\frac{1}{2}\tr\left(\begin{pmatrix}\frac{1}{N_1}\bm{J}_{N_1}&&&0\\
&\frac{1}{N_2}\bm{J}_{N_2}&&\\
&&\ddots&\\
0&&&\frac{1}{N_G}\bm{J}_{N_G}\end{pmatrix}\bm{\Delta}^2\right)\nonumber\\
&=&\frac{1}{2N}\sum_{i=1}^N\sum_{j=1}^N d^2(y_i,y_j)-\frac{1}{2}\sum_{g=1}^G \frac{1}{N_g}\sum_{i=1}^N\sum_{j=1}^N d^2(y_i,y_j)I_{gi}I_{gj}\nonumber\\
&=&T_{\bm{\Delta}}-W_{\bm{\Delta}},\nonumber
\end{eqnarray}
and since $B_{\bm{\Delta}}=T_{\bm{\Delta}}-W_{\bm{\Delta}}$, we have that $B_{\bm{\Delta}}=\tr(\bm{H}_c\bm{G}_{\bm{\Delta}})$. Also, since $W_{\bm{\Delta}}=T_{\bm{\Delta}}-B_{\bm{\Delta}}$, we find that $W_{\bm{\Delta}}=\tr\left((\bm{I}_N - \bm{H}_c)\bm{G}_{\bm{\Delta}}\right)$.

\subsection{The DBF test statistic}\label{dbf_sec}



Making use of the distance-based variance decomposition above, a distance-based test statistic has been defined to test the null hypothesis of equality between groups with respect to the chosen distance measure. This statistic, denoted DBF, is of the form
\begin{equation}\label{f_stat_0}
F_{\bm{\Delta}}=\frac{B_{\bm{\Delta}}}{W_{\bm{\Delta}}} = \frac{\tr(\bm{H}_c\bm{G}_{\bm{\Delta}})}{\tr\left((\bm{I}_N - \bm{H}_c)\bm{G}_{\bm{\Delta}}\right)},
\end{equation}
and was originally used to compare functional data arising in genomic experiments \citep{minas2011distance}. Analogously to the Dempster trace criterion and Lawley-Hotelling trace statistic, this statistic considers a ratio of between- to within-group variability. Larger values of this statistic provide evidence against the null hypothesis, as larger between-group variability and smaller within-group variability suggest that observations in the same group are more similar than observations in different groups. A statistic of similar form was also proposed by \cite{anderson2001new} for application in ecology, but with degrees of freedom divisors $G-1$ and $N-G$ in the numerator and denominator, respectively. 

When the observations are $P$-dimensional vectors, it has been shown that $F_{\bm{\Delta}}$ is related monotonically to the Lawley-Hotelling and Pillai trace MANOVA statistics for $G>2$ \citep{minas2011distance}. For instance, on defining the distance matrices $\bm{\Delta}_W=\{d_{W}(\bm{y}_i,\bm{y}_j)\}_{i,j=1}^N$ with $d^2_{W}(\bm{y}_i,\bm{y}_j)=\left(\bm{y}_i-\bm{y}_j\right)^T\bm{W}^{-1}\left(\bm{y}_i-\bm{y}_j\right)$ and $\bm{\Delta}_T=\{d_{T}(\bm{y}_i,\bm{y}_j)\}_{i,j=1}^N$ with $d^2_{T}(\bm{y}_i,\bm{y}_j)=\left(\bm{y}_i-\bm{y}_j\right)^T\bm{T}^{-1}\left(\bm{y}_i-\bm{y}_j\right)$, it was shown that
\begin{equation} 
\textrm{Lawley-Hotelling}=PF_{\bm{\Delta}_W}\quad \textrm{and}\quad \textrm{Pillai trace}=\frac{PF_{\bm{\Delta}_T}}{1+(1-P)F_{\bm{\Delta}_T}}.\nonumber
\end{equation} 
For $G=2$, it was also shown that $F_{\bm{\Delta}}$ is monotonically related  to Hotelling's $T^2$ statistic via the equation
\begin{equation}\label{t2_f}
T^2=\frac{(N-2)PF_{\bm{\Delta}_{T}}}{1+(1-P)F_{\bm{\Delta}_{T}}}.
\end{equation}  
Expanding on this relationship between $T^2$ and $F_{\bm{\Delta}_{T}}$, since we know that $(N-P-1)T^2/((N-2)P)$ has an exact F distribution under the null, it follows that  
\begin{equation}
\frac{(N-P-1)F_{\bm{\Delta}_{T}}}{1+(1-P)F_{\bm{\Delta}_{T}}}\sim F_{P,N-P-1}.\nonumber
\end{equation}
That is, a transformation of $F_{\bm{\Delta}_{T}}$ follows the exact $F$ distribution with degrees of freedom $P$ and $N-P-1$. As a special case, when the data consists of scalar observations and the Euclidean distance $d_E,$ is applied, $F_{\bm{\Delta}_E}$ is identical to the classical one-way ANOVA F statistic, ignoring the degrees of freedom divisors $G-1$ and $N-G$ in the numerator and denominator, respectively. Thus, 
\begin{equation}
F_{\bm{\Delta}}\left(\frac{N-G}{G-1}\right)\sim F_{G-1,N-G}\nonumber
\end{equation}
\citep{anderson2001new,minas2011distance}.

Given an observed value of the test statistic, $\hat{F}_{\bm{\Delta}}$, computed for any suitably chosen distance measure $d$, inference can be carried out using a non-parametric approach. That is, the p-value can be found using permutations. Given $N_{\pi}$ permutations $\pi\in\Pi$, where $\pi$ is a one-to-one mapping of the set $\{1,\ldots,N\}$ to itself, the set $\{\hat{F}_{\bm{\Delta}_{\pi}}\}_{\pi\in\Pi}$ is generated by recalculating $B_{\bm{\Delta}}$ for each permutation, denoted $\hat{B}_{\bm{\Delta}_{\pi}}$, and using the monotonic relationship 
\begin{equation}\label{f_b}
\hat{F}_{\bm{\Delta}_{\pi}}=\frac{\hat{B}_{\bm{\Delta}_{\pi}}}{T_{\bm{\Delta}}-\hat{B}_{\bm{\Delta}_{\pi}}}.
\end{equation}
${T}_{\bm{\Delta}}$ is fixed for all permutations so that permutated values of $F_{\bm{\Delta}}$ are monotonically related to permuted values
of $B_{\bm{\Delta}}$. The p-value is then computed as the proportion of the $N_{\pi}$ permuted statistics greater than or equal to the observed $\hat{F}_{\bm{\Delta}}$, i.e., 
\begin{equation}
\textrm{p-value}=\frac{\#\hat{F}_{\bm{\Delta}_{\pi}}\geq \hat{F}_{\bm{\Delta}} }{N_{\pi}}.\nonumber
\end{equation}
Clearly, this is a one-sided test, since only larger values of $F_{\bm{\Delta}}$ provide evidence against the null.

As an alternative to this expensive permutation-based testing approach, we propose an approximate distribution for the null sampling distribution of $F_{\bm{\Delta}}$ in Section \ref{db_bg_var}. From this approximate distribution p-values can be well-approximated without the need for permutations. Since $F_{\bm{\Delta}}$ is related to $B_{\bm{\Delta}}$ via \eqref{f_b}, we first approximate the null distribution of $B_{\bm{\Delta}}$. 

\section{Distance-based between-group variability}\label{db_bg_var}

\subsection{Skewed characteristics}\label{B_skewness}

For general data structures and distance measures, the null sampling distribution of DBF test statistic \eqref{f_stat_0} is unknown. The reason for this is that the between-group variability quantity, $B_{\bm{\Delta}}$, featuring in the statistic will, in general, follow some unknown distribution which depends on the specific distance measure being used \citep{mantel1967detection}. On denoting the $(i,j)^{\textrm{th}}$ element of $\bm{H}_c$ by $h_{ij}$ and recalling that $\bm{H}_c$ is centered, $B_{\bm{\Delta}}$ can be expressed as the weighted sum of squared distances
\begin{equation}
B_{\bm{\Delta}}=-\frac{1}{2}\sum_{i\neq j}^N h_{ij}d^2(y_i,y_j).\nonumber
\end{equation}
Thus even if each $d^2(y_i,y_j)$, for $i\neq j$, was assumed to be a random variable with known distribution, $B_{\bm{\Delta}}$ would be a weighted sum of correlated and uncorrelated random variables, whose distribution would be difficult to evaluate. For instance, the problem of evaluating the sum of correlated and uncorrelated Chi-squared and Gamma random variables has been considered extensively (see, for example, \cite{solomon1977distribution}, \cite{kourouklis1985distribution}). Although it has been argued that $B_{\bm{\Delta}}$ has the appearance of a U-statistic which is asymptotically normal \citep{mantel1967detection,hoeffding1948class}, in our experience with different vectorial and non-vectorial data types arising in real applications, even for large samples sizes, $B_{\bm{\Delta}}$ often appears to be skewed to various degrees. 

Further insights into this problem can be obtained by exploring the empirical permutation distribution of $B_{\bm{\Delta}}$ for a number of real datasets involving different data structures and distances. As an illustration, we consider three different examples: 

\begin{enumerate}[(i)]
\item Vectorial and real-valued data: the data consists of the $P=50$ gene expression measurements observed on $N=103$ biological samples from the Novartis multi-tissue dataset described in \cite{monti2003consensus}. In this case, $G=4$, corresponding to four different tissues. For this dataset, we considered the Euclidean, Mahalanobis and Manhattan distances. 
\item Vectorial and discrete-valued data: the data consists of $P=5$ randomly selected SNPs observed on $N=254$ samples from the ADNI dataset (see Section \ref{adni_gwas} for further details). The observation of each sample at each SNP is the number of minor alleles,  taking one value in $\{0,1,2\}$. In this case, $G=2$, corresponding to the two groups being compared, healthy controls and Alzheimer's disease patients. Here we used the identity-by-state (IBS), Rogers and Tanimoto I, and Sokal and Sneath genetic distances. The IBS distance compares the number of minor alleles at each SNP in the set of SNPs, while the Rogers and Tanimoto I and Sokal and Sneath distances use a function of the total number of matches of minor alleles across the whole set of SNPs. See Appendix for further details.
\item Functional data (curves): the data consists of $N=18$ gene expression functional data replicates for a randomly selected gene in a dataset on {\it M.tuberculosis} analyzed by \cite{tailleux2008probing}. In this case, $G=2$, corresponding to two different types of cell, and replicate time courses were observed at $4$ time-points for each type of cell. These were smoothed via cubic smoothing splines to yield the $18$ replicate curves \citep{minas2011distance}. Figure \ref{m_tub_gene_curves} shows observed time courses and their fitted curves for two randomly selected genes. The L$_2$, Visual L$_2$ and Curvature distance measures were applied to this dataset. The L$_2$ measure captures the difference in magnitude between curves, the Visual L$_2$  measure captures their scale-invariant differences in shape, and the Curvature measure captures their difference in rate of change regardless of direction. See Appendix for further details. 

\begin{figure*}[h!]
\center
\includegraphics[scale=0.64]{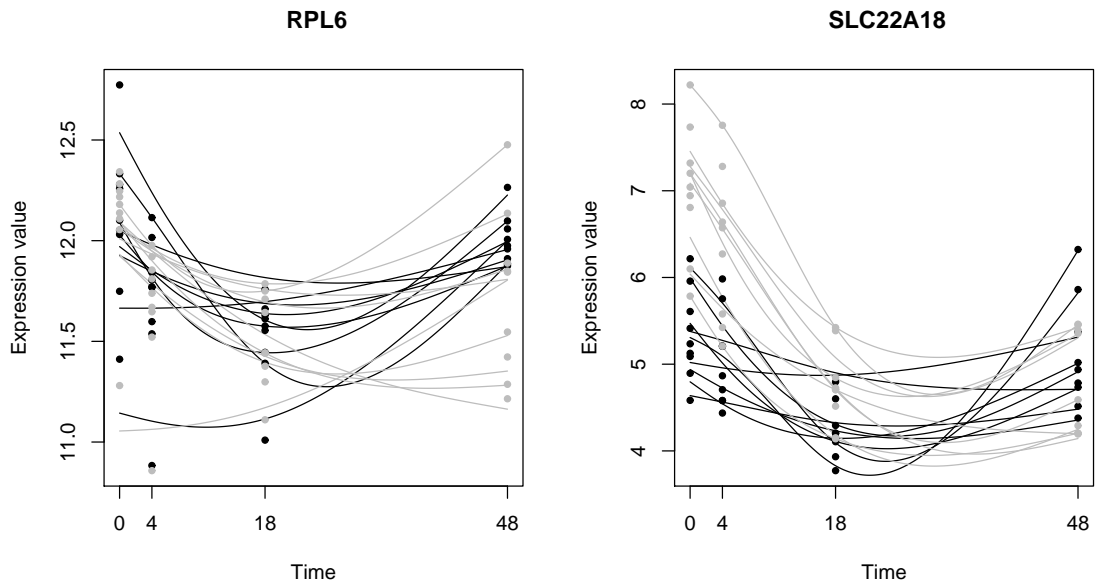}
\caption{Replicate gene expression time courses modeled as time-dependent curves for genes RPL6 and SLC22A18 of the {\it M. tuberculosis} dataset; black for dendritic cells and grey for macrophages. The points represent the original gene expression time course measurements.}
\label{m_tub_gene_curves}
\end{figure*}

\end{enumerate}

\begin{figure*}[h!]
\center
\includegraphics[scale=0.735]{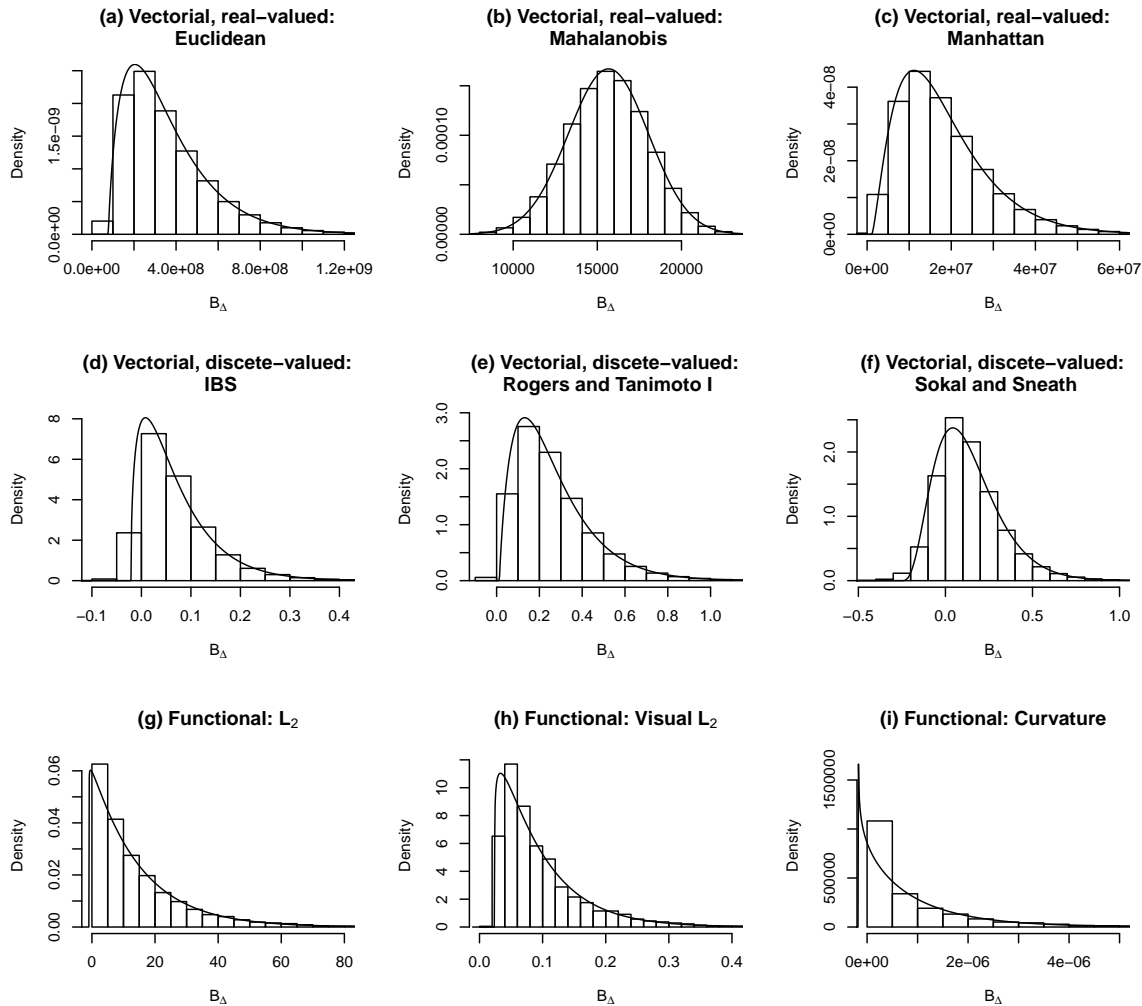}
\caption{Sampling distributions of $B_{\bm{\Delta}}$ obtained by using $10^6$ Monte Carlo permutations for three different data types and corresponding distances. (a)-(c) Vectorial and real-valued gene expression data with $N=103$. (d)-(f) Vectorial and discrete-valued SNP data with $N=254$. (g)-(i) Functional representation of longitudinal gene expression data with $N=18$. Overlayed is the proposed approximate probability density function described in Section \ref{pearson}.}
\label{tr_hg_hists}
\end{figure*}

The exact permutation distribution of $B_{\bm{\Delta}}$ in each case would be given by the set $\{\hat{B}_{\bm{\Delta}_{\pi}}\}_{\pi\in\Pi}$ where $\Pi$ contains all $N!$ permutations $\pi$ of the elements of $\{1,\ldots,N\}$. For large $N$, due to the computational effort required in enumerating all possible permutations, the exact distribution is generally unavailable. Figure \ref{tr_hg_hists} shows the approximate sampling distribution of $B_{\bm{\Delta}}$ obtained by using $10^6$ Monte Carlo permutations for each of the data types and distance measures considered. In almost all cases the distributions are heavily skewed. Remarkably, this is also the case for large sample sizes.

\subsection{A Person Type III approximation} \label{pearson}

In order to approximate the exact permutation distribution $B_{\bm{\Delta}}$, we propose an approach based on moment matching. Given the skewness often observed in real datasets, we assume that $B_{\bm{\Delta}}$ follows a Pearson type III distribution. This is a type of Gamma distribution encompassing other distributions as special cases, such as the Exponential and Normal distributions \citep{josse2008testing}, and has been successfully used for modeling skewed sampling distributions of various other statistics \citep{berry1983moment, kazi1995refined, josse2008testing}.
  

The first three moments of the exact permutation distribution of $B_{\bm{\Delta}}$ are given by
\begin{equation}
\mu_B=\frac{1}{N!}\sum_{\pi\in\Pi}\hat{B}_{\bm{\Delta}_{\pi}},\quad\sigma^2_B=\frac{1}{N!}\sum_{\pi\in\Pi}\hat{B}^2_{\bm{\Delta}_{\pi}}-{\mu}^2_B, \quad \text{ and } \quad
\gamma_B=\frac{\frac{1}{N!}\sum_{\pi\in\Pi}\hat{B}^3_{\bm{\Delta}_{\pi}}-\mu^2_B\sigma^2_B-\mu^3_B}{\sigma^3_B},\nonumber
\end{equation}
respectively. By evaluating the expressions analytically, closed form manipulations of the three moments have been derived allowing their efficient computation for $N>6$ without the need for permutations \citep{kazi1995refined}. These closed form expressions require only the square, symmetric and centered matrices $\bm{H}_c$ and $\bm{G}_{\bm{\Delta}}$. The mean and variance, for example, are given by
\begin{eqnarray}
\mu_B=\frac{A_1B_1}{N-1}\quad\textrm{and}\quad\sigma^2_B&=&\frac{2\left(\left(N-1\right)A_2-A_1^2\right)\left(\left(N-1\right)B_2-B_1^2\right)}{(N-1)^2(N+1)(N-2)}\nonumber\\
&&+\left[\frac{\left(N(N+1)A_3-(N-1)\left(A_1^2+2A_2\right)\right)}{(N+1)N(N-1)(N-2)(N-3)}\right.\nonumber\\
&&\left.\times \left(N(N+1)B_3-(N-1)\left(B_1^2+2B_2\right)\right)\right],\nonumber
\end{eqnarray}
where $A_1=\tr\left(\bm{H}_c\right)$, $A_2=\tr\left(\bm{H}_c\bm{H}_c\right)$, $A_3=\tr\left(\bm{H}_c^2\right)$, $B_1=\tr\left(\bm{G}_{\bm{\Delta}}\right)$, $B_2=\tr\left(\bm{G}_{\bm{\Delta}}\bm{G}_{\bm{\Delta}}\right)$ and 
$B_3=\tr\left(\bm{G}_{\bm{\Delta}}^2\right)$, where, for matrix $\bm{A}=\{a_{ij}\}_{i,j=1}^N$, $\bm{A}^2=\{a_{ij}^2\}_{i,j=1}^N$. The expression for skewness is much more involved; the reader is referred to \cite{kazi1995refined}.

On standardizing $B_{\bm{\Delta}}$ by subtracting $\mu_B$ and dividing by $\sigma_B$, the Pearson type III distribution can be parameterized by only the skewness parameter, $\gamma_B$. That is,
\begin{equation}
B^s_{\bm{\Delta}}=\frac{B_{\bm{\Delta}}-\mu_B}{\sigma_B}\sim PT_{III}\left(\gamma_B\right),\nonumber
\end{equation}
where $PT_{III}$ denotes the Pearson type III distribution. By assumption of this  model, the support of random variable $B^s_{\bm{\Delta}}$ is given by $[-2/\gamma_B,\infty)$ if $\gamma_B>0$, and $(-\infty, -2/\gamma_B]$ if $\gamma_B<0$, and we denote the cumulative distribution function (CDF) of $B^s_{\bm{\Delta}}$ by $\mathcal{F}_{B^s_{\bm{\Delta}}}(b;\gamma_B)$, and probability density function (PDF) of $B^s_{\bm{\Delta}}$ by $f_{B^s_{\bm{\Delta}}}(b;\gamma_B)$. The PDF $f_{B^s_{\bm{\Delta}}}(b;\gamma_B)$ is defined by 
\begin{equation}
\frac{\left(2/\gamma_B\right)^{4/\gamma_B^2}}{\Gamma\left(4/\gamma_B^2\right)}\left(\frac{2+\gamma_B b}{\gamma_B}\right)^{(4-\gamma_B^2)/\gamma_B^2} \exp\left(-\frac{2(2+\gamma_B b)}{\gamma_B^2}\right)\nonumber
\end{equation}
for $\gamma_B>0$ and $-2/\gamma_B\leq b<\infty$, where $\Gamma(\cdot)$ denotes the usual Gamma function,  
\begin{equation}
\frac{\left(-2/\gamma_B\right)^{4/\gamma_B^2}}{\Gamma\left(4/\gamma_B^2\right)}\left(\frac{-(2+\gamma_B b)}{\gamma_B}\right)^{(4-\gamma_B^2)/\gamma_B^2} \exp\left(\frac{-2(2+\gamma_B b)}{\gamma_B^2}\right)\nonumber
\end{equation}
for $\gamma_B<0$ and $-\infty<b\leq-2/\gamma_B$, and the standard Normal distribution for $\gamma_B=0$ \citep{mielke2007permutation}.

In practice $\gamma_B$ is not equal to zero exactly, as exhibited for some real datasets in Figure \ref{tr_hg_hists}. We see that all sampling distributions, and corresponding approximate PDFs, are skewed to some degree. Thus in the next section we only consider the cases where $\gamma_B>0$ and $\gamma_B<0$, and do not consider the trivial case of $\gamma_B=0$.


\section{Approximate null distribution of the DBF statistic}\label{null_dbf_dist}

The DBF statistic $F_{\bm{\Delta}}$ is related to the standardized distance-based between-group variability quantity $B^s_{\bm{\Delta}}$ via the one-to-one function $h:B^s_{\bm{\Delta}}\mapsto F_{\bm{\Delta}}$ defined by
\begin{equation}\label{f_t0}
h\left(B^s_{\bm{\Delta}}\right)=\frac{\mu_B+\sigma_B B^s_{\bm{\Delta}}}{T_{\bm{\Delta}}-\mu_B-\sigma_B B^s_{\bm{\Delta}}},
\end{equation}
with inverse $h^{-1}: F_{\bm{\Delta}}\mapsto B^s_{\bm{\Delta}}$ defined by
\begin{equation}\label{f_t0_inv}
h^{-1}\left(F_{\bm{\Delta}}\right)=\frac{\left(T_{\bm{\Delta}}-\mu_B\right)F_{\bm{\Delta}}-\mu_B}{\sigma_B\left(1+F_{\bm{\Delta}}\right)}.
\end{equation}

\noindent We aim to derive the approximate null distribution of $F_{\bm{\Delta}}$ in terms of the distribution of $B^s_{\bm{\Delta}}$ via transformation $h$, which is required to be continuous  over the given supports of $B^s_{\bm{\Delta}}$. However, it is not continuous in the positive plane at $\beta=(T_{\bm{\Delta}}-\mu_B)/\sigma_B$ because  $T_{\bm{\Delta}}=\tr\left(\bm{G}_{\bm{\Delta}}\right)>\mu_B$ due to $\tr(\bm{H}_c)=1$. Since the boundaries of both supports of $B^s_{\bm{\Delta}}$ depend on the skewness, the position of $\beta$ may or may not affect the continuity of the distribution of $B^s_{\bm{\Delta}}$ over the given support. We thus consider dealing with the discontinuity separately for both supports of $B^s_{\bm{\Delta}}$, i.e. for both cases of skewness.

\begin{figure*}[h!]
\center
\includegraphics[scale=0.65]{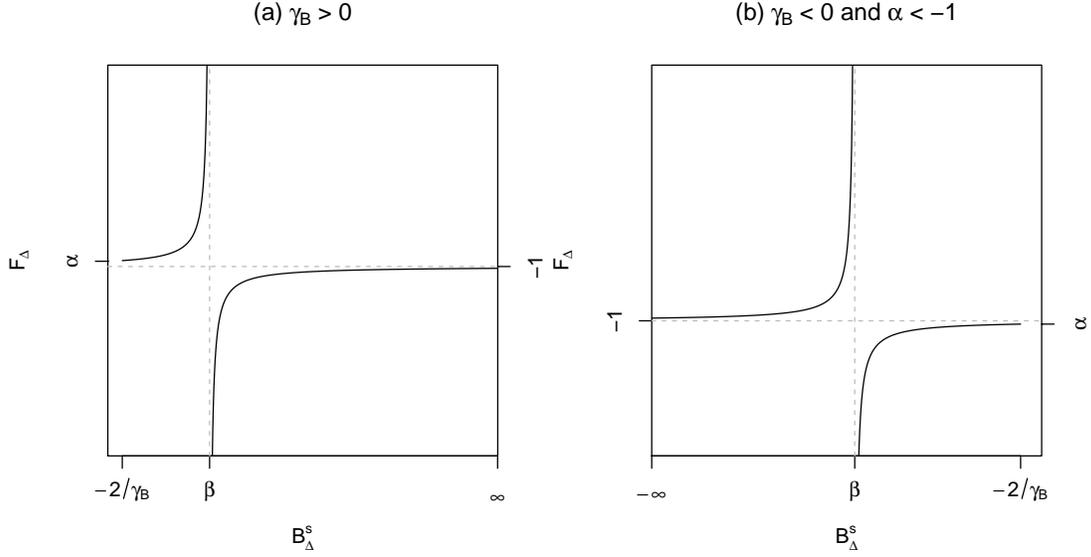}
\caption{(a) $F_{\bm{\Delta}}$ and $B^s_{\bm{\Delta}}$ are monotonically related everywhere except at $B^s_{\bm{\Delta}}=\beta$ for ${\gamma}_B>0$ over the support $[-2/{\gamma}_B,\infty)$. (b) $F_{\bm{\Delta}}$ and $B^s_{\bm{\Delta}}$ are monotonically related everywhere except at $B^s_{\bm{\Delta}}=\beta$ for ${\gamma}_B<0$ over the support $(-\infty,-2/{\gamma}_B]$ when $\alpha < -1$.}
\label{all_skew_alpha}
\end{figure*}

First consider the positive skewness case, where the support of $B^s_{\bm{\Delta}}$ is $[-2/\gamma_B,\infty)$. Since $\gamma_B>0$, $-2/\gamma_B$ is negative and the discontinuity shows itself as $B^s_{\bm{\Delta}}$ increases form $-2/\gamma_B$ to $\infty$. Figure \ref{all_skew_alpha} (a) shows how $F_{\bm{\Delta}}$ behaves as a function of $B^s_{\bm{\Delta}}$ over this support; $F_{\bm{\Delta}}$ is an increasing function of $B^s_{\bm{\Delta}}$ on both sides of the discontinuity at $\beta$. 

We thus divide the support of $B^s_{\bm{\Delta}}$ into the two regions, namely 
\begin{equation}
\left[\frac{-2}{\gamma_B},\beta\right)\quad\textrm{and}\quad \left(\beta,\infty\right),\nonumber
\end{equation}
where the equivalent supports of $F_{\bm{\Delta}}$ are given by $[\alpha,\infty)$, where 
\begin{equation}\label{f_a}
\alpha=\frac{\gamma_B\mu_B-2\sigma_B}{\gamma_B\left(T_{\bm{\Delta}}-\mu_B\right)+2\sigma_B}
\end{equation}
satisfies $h\left(\alpha\right)=-2/\gamma_B$, and $(-\infty,-1)$. We can show by contradiction that $\alpha > -1$, since $\alpha \leq -1$ implies $T_{\bm{\Delta}}\gamma_B\leq 0$, and by definition both $T_{\bm{\Delta}}$ and $\gamma_B$ are positive. In these regions we can apply the transformation  since there are no discontinuities, and define the CDF of $F_{\bm{\Delta}}$ in terms of the CDF of $B^s_{\bm{\Delta}}$  by
\begin{equation}\label{CDF_f_pos_skew}
\mathcal{F}_{F_{\bm{\Delta}}}\left(f;\mu_B,\sigma_B,\gamma_B\right)=\begin{cases}
\mathcal{F}_{B^s_{\bm{\Delta}}}\left( h^{-1}(f);\gamma_B\right)-\mathcal{F}_{B^s_{\bm{\Delta}}}\left(\beta;\gamma_B\right)& -\infty<f<-1 \\
1+\mathcal{F}_{B^s_{\bm{\Delta}}}\left( h^{-1}(f);\gamma_B\right)-\mathcal{F}_{B^s_{\bm{\Delta}}}\left(\beta;\gamma_B\right)& \alpha\leq f<\infty
\end{cases}
\end{equation}
for $\gamma_B>0$. The relevant derivations are in the Appendix, including a proof that this is a valid CDF. 

Now we turn our attention to the negative skewness case, where the support of $B^s_{\Delta}$ is $(-\infty,-2/{\gamma}_B]$. In this case, ${\gamma}_B<0$ so that $-2/{\gamma}_B$ is positive. The discontinuity at $B^s_{\bm{\Delta}}=\beta$ thus only needs to be considered if $\beta$ is to the left of $-2/{\gamma}_B$, otherwise it can be ignored since it is not included in the support of $B^s_{\bm{\Delta}}$. We consider these two cases separately. First consider the case where $\beta<-2/{\gamma}_B$. We have that 
\begin{equation}
\beta<\frac{-2}{{\gamma}_B}\Rightarrow \frac{-{\mu}_B}{{\sigma}_B}<\frac{T_{\bm{\Delta}}-{\mu}_B}{{\sigma}_B}<\frac{-2}{{\gamma}_B},\nonumber
\end{equation}
since $T_{\bm{\Delta}}>{\mu}_B$, from which we find
\begin{equation}
0<\frac{{\gamma}_B T_{\bm{\Delta}}}{{\gamma}_B{\mu}_B-2{\sigma}_B}<1.
\end{equation}
Applying this with the equation for $\alpha$ given by \eqref{f_a} yields $\alpha < -1$. Thus we define the occurrence of this first case when $\alpha<-1$. Figure \ref{all_skew_alpha} (b) shows how $F_{\bm{\Delta}}$ behaves as a function of $B^s_{\bm{\Delta}}$ over the support $(-\infty,-2/{\gamma}_B]$ when $\alpha<-1$. As with the positive skewness case, $F_{\bm{\Delta}}$ is an increasing function of $B^s_{\bm{\Delta}}$ on both sides of the discontinuity at $\beta$. We thus divide the support of $B^s_{\bm{\Delta}}$ into the two regions 
\begin{equation}
\left(-\infty,\beta\right)\quad\textrm{and}\quad \left(\beta,\frac{-2}{{\gamma}_B}\right],\nonumber
\end{equation}
with equivalent supports of $F_{\bm{\Delta}}$ given by $(-1,\infty)$ and $(-\infty,\alpha]$, in which $F_{\bm{\Delta}}$ and $B^s_{\bm{\Delta}}$ are monotonically related with no discontinuities. Thus we define the CDF of $F_{\bm{\Delta}}$ as 
\begin{equation}\label{CDF_f_neg_skew_1}
\mathcal{F}_{F_{\bm{\Delta}}}\left(f;{\mu}_B,{\sigma}_B,{\gamma}_B\right)=\begin{cases}
\mathcal{F}_{B^s_{\bm{\Delta}}}\left( h^{-1}(f);{\gamma}_B\right)-\mathcal{F}_{B^s_{\bm{\Delta}}}\left(\beta;{\gamma}_B\right)& -\infty<f\leq \alpha \\
1+\mathcal{F}_{B^s_{\bm{\Delta}}}\left( h^{-1}(f);{\gamma}_B\right)-\mathcal{F}_{B^s_{\bm{\Delta}}}\left(\beta;{\gamma}_B\right)& -1 < f<\infty
\end{cases}
\end{equation}
for $\gamma_B<0$ and $\alpha< -1$.



Now consider the case where $\beta>-2/{\gamma}_B$; this is equivalent to $\alpha>-1$. In this case there are no discontinuities in the support of $B^s_{\bm{\Delta}}$, so $F_{\bm{\Delta}}$ and $B^s_{\bm{\Delta}}$ are monotonically related everywhere. The support for $B^s_{\bm{\Delta}}$ given by $(-\infty,-2/{\gamma}_B]$ is equivalent to the support for $F_{\bm{\Delta}}$ of $(-1,\alpha]$. Thus the CDF of $F_{\bm{\Delta}}$ is defined as
\begin{equation}\label{CDF_f_neg_skew_2}
\mathcal{F}_{F_{\bm{\Delta}}}\left(f;{\mu}_B,{\sigma}_B,{\gamma}_B\right)=\begin{cases}
0& -\infty < f <-1 \\
\mathcal{F}_{B^s_{\bm{\Delta}}}\left( h^{-1}(f);{\gamma}_B\right)& -1 < f\leq\alpha\\
1&\alpha<f<\infty
\end{cases}
\end{equation}
for ${\gamma}_B<0$ and $\alpha > -1$. 

Using these results, the approximate p-value of an observed $\hat{F}_{\bm{\Delta}}$ can be readily obtained without permutations. On computing the permutational mean ${\mu}_{B}$, variance ${\sigma}^2_B$, and skewness ${\gamma}_B$, and additionally $\alpha$ if ${\gamma}_B<0$, the p-value is given by $1-\mathcal{F}_{F_{\bm{\Delta}}}\left(\hat{F}_{\bm{\Delta}};{\mu}_{T},{\sigma}_B,{\gamma}_B\right)$, where $\mathcal{F}_{F_{\bm{\Delta}}}\left(\cdot;{\mu}_{T},{\sigma}_B,{\gamma}_B\right)$ is the CDF chosen for the specific case of skewness and $\alpha$ value. 



For the given case of skewness and $\alpha$ value, the PDF of $F_{\bm{\Delta}}$, denoted $f_{F_{\bm{\Delta}}}\left(f;{\mu}_{T},{\sigma}_B,{\gamma}_B\right)$, is given in terms of $f_{B^s_{\bm{\Delta}}}(b;{\gamma}_B)$ by differentiating the CDF. Thus we have that 
\begin{eqnarray}
f_{F_{\bm{\Delta}}}\left(f;{\mu}_{T},{\sigma}_B,{\gamma}_B\right)&=&\left|\frac{d}{df}h^{-1}(f)\right|f_{B^s_{\bm{\Delta}}}(h^{-1}(f);{\gamma}_B)\nonumber\\
&=&\frac{T_{\bm{\Delta}}}{{\sigma}_B(1+f)^2}f_{B^s_{\bm{\Delta}}}(h^{-1}(f);{\gamma}_B),\nonumber
\end{eqnarray}
where the range of $f$ is given by the selected case of CDF.

\section{Simulation Results}\label{sims}

\subsection{Comparison with ANOVA and MANOVA}

Given that $F_{\bm{\Delta}}$ equals the ANOVA F statistic up to a constant as a special case for univariate data, we verify that the distribution of $F_{\bm{\Delta}}$ approximates that of the ANOVA F statistic well as $N$ and $G$ increase. Also, since $F_{\bm{\Delta}}$ is related to Hotelling's $T^2$ as a special case for multivariate data with $G=2$, we verify that our proposed distribution, transformed via $\eqref{t2_f}$, approximates the distribution of $T^2$ well as $N$ increases. That is, we aim to show that for the special cases the DBF test is approximately equivalent to the ANOVA F and Hotelling's $T^2$ tests, respectively.

For the univariate case, data was generated under the null and the DBF statistic using the Euclidean distance measure, $d_E$, and the ANOVA F statistic were computed. P-values were found by comparing against their respective distributions. For $N=40,100,500,1000$ and $G=2,4,5$, the $k^{\text{th}}$ Monte Carlo run consisted of simulating $y_1,\ldots,y_N \sim N(\mu_k,\sigma_k^2)$, where $\mu_k\sim U(-10,10)$, $\sigma_k^2\sim U(0,10)$, and $U(a,b)$ denotes the Uniform distribution over $[a,b]$. The mean and standard deviation of the absolute differences between the p-values obtained for $B=200$ Monte Carlo simulations are reported in Table \ref{ANOVA_1}. It can be seen that as $N$ and $G$ increase, the absolute difference between the p-values decreases, thus showing that the approximate distribution of the DBF statistic behaves as expected in this case.

%
%

\begin{table*}[h!]
\begin{center} 
\caption{Mean (and standard deviation) of the absolute differences between p-values of the DBF statistic, and ANOVA F ($P=1$) and Hotelling's $T^2$ ($P=10$) statistics, under the null for $200$ Monte Carlo runs.}\label{ANOVA_1}

\begin{tabular}{l|ll|ll|ll|ll}
        \toprule
& \multicolumn{6}{c}{$P=1$}&\multicolumn{2}{c}{$P=10$}  \\
$N$ & \multicolumn{2}{c}{$G=2$} & \multicolumn{2}{c}{$G=4$}  &\multicolumn{2}{c}{$G=5$} &\multicolumn{2}{c}{$G=2$}   \\
\midrule
$40$& $0.0252$& $(0.0342)$ & $0.00670$& $(0.00535)$& $0.00565$& $(0.00464)$&$0.004702$& $(0.00272)$\\ 
$100$&$0.0137$& $(0.0225)$ &$0.00306$& $(0.00267)$&$0.00254$ &$(0.00196)$&$0.00200$& $(0.00121)$ \\
$500$&$0.00474$ &$(0.00951)$ &$0.000594$& $(0.000525)$&$0.000541$& $(0.000383)$&$0.000392$& $(0.000260)$\\
$1000$&$0.00276$& $(0.00599)$ &$0.000344$& $(0.000272)$&$0.000278$& $(0.000196)$&$0.000212$& $(0.000132)$\\
\bottomrule
\end{tabular}
\end{center}

\end{table*}

For the multivariate case, the DBF statistic using the Mahalanobis-like distance measure $d_T$, and the Hotelling's $T^2$ statistic were computed. P-values were found by comparing against their respective distributions. For $N=40,100,500,1000$ and $P=10$, the $k^{\text{th}}$ Monte Carlo run consisted of simulating $\bm{y}_1,\ldots\bm{y}_N\sim N_P(\bm{\mu}_k,\bm{\Sigma_k})$, where $\bm{\mu}_k=(\mu_{k1},\ldots,\mu_{Pk})^T$ with $\mu_{jk}\sim U(-6,6)$ for $j=1,\ldots,P$, and $\bm{\Sigma}_k$ a random Wishart matrix of size $P\times P$. The mean and standard deviation of the absolute differences between the p-values obtained for $B=200$ Monte Carlo runs are reported in Table \ref{ANOVA_1}. As $N$ increases the difference between the p-values decreases, showing that the DBF and Hotelling's $T^2$ tests are approximately equivalent as $N$ increases.

A further experiment was performed to show that, as $N$ increases, the proposed approximate null distribution of the DBF statistic approximates the true ANOVA F and Hotelling's $T^2$ distributions, on applying the required transformations, quite well. In particular, we show that it yields a better approximation than a permutation-based CDF, especially when the number of permutations is low. 

\begin{figure*}[h!]
\center
\includegraphics[scale=0.73]{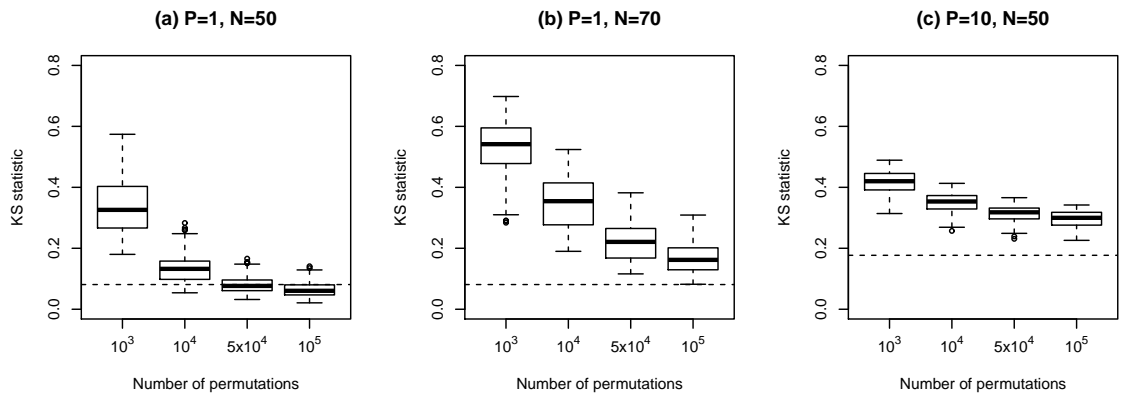}
\caption{(a)-(b) Empirical distributions of the KS statistic quantifying the distance between the DBF permutation CDF, suitably transformed, and the ANOVA F CDF, for each set of Monte Carlo permutations. The dotted line represents the KS statistic comparing the approximate DBF CDF, suitably transformed, and the ANOVA F CDF. (c) Empirical distributions of the KS statistic quantifying the distance between the DBF permutation CDF, suitably transformed, and the Hotelling's $T^2$ CDF, for each set of Monte Carlo permutations. The dotted line represents the KS statistic comparing the approximate DBF CDF, suitably transformed, and the Hotelling's $T^2$ CDF.}
\label{boxplots1}
\end{figure*}

For $P=1$, $G=2$, and each of $N=50,70$, one set of univariate observations were generated under the null from a Normal distribution as above. The DBF null CDF, suitably transformed, and the ANOVA F CDF were obtained, and the Kolmogorov-Smirnov (KS) statistic was used to compute the difference between these distributions. This statistic is computed as the maximum distance between two vectors representing the CDFs of interest; we used a vector of $1000$ equally spaced points across the range of the approximate DBF distribution. For the given dataset for each $N$, and for each of $B=200$ Monte Carlo runs, we used an increasing set of Monte Carlo permutations to compute the permutation CDF of the DBF statistic. We used $10^3, 10^4, 5\times 10^4$ and $10^5$ permutations, so that for each Monte Carlo run, $10^3$ Monte Carlo permutations were enumerated, then $9\times 10^3$ Monte Carlo permutations were added to yield the larger set of $10^4$ permutations and so on. For each of these $4$ sets of permutations the KS statistic depicting the difference between the DBF permutation CDF and the ANOVA F CDF was computed. This yielded an empirical distribution of $200$ KS statistic values for each set of permutations. The results of this experiment are shown in plots (a) and (b) of Figure \ref{boxplots1}. We see that for $N=50$, using more than $5\times 10^4$ permutations yields a permutation distribution which is directly comparable with our approximate distribution. For $N=70$, however, the approximate DBF distribution better approximates the true underlying ANOVA F distribution than the permutation distributions typically used in practice; typically not more than $10^5$ permutations are used for real data analyses.

For $P=10$, $G=2$ and $N=50$, one set of multivariate observations were generated under the null from a Multivariate Normal distribution as described above.  The DBF null CDF, suitably transformed, and the Hotelling's $T^2$ CDF were obtained. Repeating as above, and using the KS statistic to quantify the difference between the transformed DBF permutation CDF and true Hotelling's $T^2$ CDF, the results are given in plot (c) of Figure \ref{boxplots1}. We see that for $N=50$ the approximate DBF distribution yields a better approximation of the true distribution than the permutation distributions.

\subsection{Approximate null distribution for various data types and distances}\label{perf1}

In this section we illustrate how the approximate distribution of the DBF statistic compares with the Monte Carlo permutation distribution for a number of data types and distances. In our setting, we explore a range of sample sizes and distance measures for datasets simulated to mimic the real datasets introduced in Section \ref{B_skewness}. For vectorial and real-valued data, we consider the Euclidean, Bray-Curtis, Canberra, Manhattan and Maximum distances. For vectorial and discrete-valued genetic data, we consider the identity-by-state (IBS), Simple Matching, Sokal and Sneath, Rogers and Tanimoto I (RTI) and Hamman I distances. For functional data we consider the L$_2$, Visual L$_2$ and Curvature distances (see Appendix for further details on the genetic and functional distance measures). On selection of data type, distance measure and number of samples $N$, the datasets are simulated as follows:

\begin{enumerate}[(i)]
\item Vectorial and real-valued data: $P$-dimensional vectors $\bm{y}_i=(y_{i1},\ldots,y_{iP})^T$ were simulated such that $y_{ij}\sim N(0,4)$ for $i=1,\ldots,N$ and $j=1,\ldots,1000$. For the Bray-Curtis and Canberra distances where positive values are required, we take absolute values. 
\item Vectorial and discrete-valued data: $5$-dimensional vectors $\bm{y}_i=(y_{i1},\ldots,y_{i5})^T$ for $i=1,\ldots,N$ were simulated based on the observations of the $153$ control subjects from chromosome $1$ of the ADNI dataset introduced in Section \ref{adni_gwas}. $N$ control subjects were randomly selected and their minor allele counts at $5$ randomly chosen SNPs across the chromosome selected. 
\item Functional data (curves): $N$ curves $\{y_i(t)\}_{i=1}^N$ were simulated over the range $t\in [0,48]$ by using quadratic Bezier curves \citep{Farin1992} and a sampling procedure involving smoothing splines \citep{Ramsay2005}. $N$ Bezier curves were randomly generated, and $N$ $1000$-dimensional vectors were sampled from these curves at equally spaced points across $[0,48]$ with standard Gaussian error. These were then smoothed via cubic smoothing splines to yield the $N$ curves; see \cite{minas2011distance} for further details of this curve-generation procedure. This procedure generates random curves similar to those observed in real longitudinal datasets, such as those shown in Figure \ref{m_tub_gene_curves}. \end{enumerate}

\begin{table*}[h!]
\begin{center} 
\caption{Mean (and standard deviation) of the absolute differences between theoretical and permutation p-values of the DBF statistic under the null with $200$ Monte Carlo runs for vectorial, SNP and functional (curve) distances. For $N=10$, all $10!$ permutations were used, and for $N=30,100$, $10^6$  Monte Carlo permutations were used.}
\label{tab:table_n_10_30_100_exact_1_mil_all_dists}
\begin{tabular}{c|c|ll|ll|ll}
        \toprule
&& \multicolumn{6}{c}{$N$}\\
$\begin{array}{c}\textrm{Data}\\\textrm{type}\end{array}$&$\begin{array}{c}\textrm{Distance}\\\textrm{measure}\end{array}$ & \multicolumn{2}{c}{$10$}&\multicolumn{2}{c}{$30$}&\multicolumn{2}{c}{$100$}\\
\midrule
&Euclidean& $0.0159$& $(0.0139)$&$0.000706$& $(0.000554)$&$0.000317$& $(0.000266)$\\
Vector,&Bray-Curtis&$0.0135$& $(0.0110)$&$0.000664$ &$(0.000566)$&$0.000297$& $(0.000259)$\\
real-&Canberra&$0.0154$& $(0.0133)$&$0.000762$& $(0.000635)$&$0.000318$& $(0.000261)$\\
valued&Manhattan&$0.0141$& $(0.0127)$&$0.000565$& $(0.000453)$&$0.000314$& $(0.000254)$\\
&Maximum&$0.0165$& $(0.0142)$&$0.000675$& $(0.000550)$&$0.000314$& $(0.000231)$\\
\midrule
&IBS& $0.0223$& $(0.0180)$&$0.00507$& $(0.00484)$&$0.00174$& $(0.00108)$\\
Vector,&Simple Matching&$0.0222$& $(0.0206)$&$0.00321$& $(0.00301)$&$0.00152$& $(0.000964)$\\
discrete-&Sokal and Sneath&$0.0217$& $(0.0187)$&$0.00551$& $(0.00509)$&$0.00393$& $(0.00220)$\\
valued&RTI&$0.0237$& $(0.0197)$&$0.00212$& $(0.00189)$&$0.000646$& $(0.000405)$\\
&Hamman I&$0.0211$& $(0.0201)$&$0.00324$& $(0.00317)$&$0.00158$& $(0.000988)$\\
\midrule
&L$_2$& $0.0267$& $(0.0256)$&$0.00870$& $(0.00803)$&$0.00595$& $(0.00573)$\\
Functional&Visual L$_2$&$0.0370$& $(0.0332)$&$0.0130$& $(0.0118)$&$0.00885$& $(0.00841)$\\
&Curvature&$ 0.0515$& $(0.0502)$&$0.0262$& $(0.0238)$&$0.00880$& $(0.00975)$\\
\bottomrule
\end{tabular}
\end{center}

\end{table*}

We compared the theoretical and permutation p-values resulting from applying the DBF test under the null for the different distances applied to each data type. For $N=10,30,100$ and $G=2$, $B=200$ Monte Carlo runs were performed, where for each run data was generated under the null, i.e., no group effect. For $N=10$, all $N!$ permutations were used to compute the permutation p-value, but for $N=30,100$, a Monte Carlo set of $10^6$ permutations was used. The theoretical and permutation p-values were computed, and the mean and standard deviation of the absolute difference between these for each combination of data type, distance measure and $N$ are reported in Table \ref{tab:table_n_10_30_100_exact_1_mil_all_dists}. As expected, the absolute difference between the p-values decreases as $N$ increases for each distance measure applied to each data type.

\section{A genome-wide association study of Alzheimer's disease}\label{adni_gwas}

In traditional case-control association studies, subjects are genotyped for a comprehensive range of genetic markers across the entire genome, and markers associated with disease risk are sought. The objective of such studies is to identify genetic variants in the human genome that are associated with disease risk (see, for example, \cite{altshuler2008genetic} and \cite{pearson2008interpret} for good overviews of traditional GWA studies).

Genetic variations are captured by observing SNPs. Human SNPs are biallelic genetic markers, and as such, the genotype of an individual at a given SNP is represented by one of three combinations of two alleles; a major allele occurring more commonly in the study cohort, and a minor allele which is less common. The possible combinations of alleles are `major, major', `major, minor' and `minor, minor', and the genotype of the individual at a given SNP is summarized by the minor allele count. The genotype of individual $i$ at a given SNP $k$, denoted $y_{ik}$, is thus represented by either a $0$, $1$ or $2$ corresponding to homozygotes for the major allele, heterozygotes and homozygotes for the minor allele, respectively.

A common approach of scanning the genome in search of causal variants is to group SNPs together into subsets where it is plausible that some dependence exists between them, for example, if the SNPs are in the same gene or biological pathway; generally referred to as the multi-locus approach \citep{mukhopadhyay2010association,wu2010powerful, yang2009genome}. Sliding windows which partition the genome into overlapping subsets of SNPs of the same length have also been proposed to group SNPs together, and ensure coverage of the entire genome without omitting possibly useful information in intergenic regions. 

We describe an application of the DBF test with the approximate inference approach of Section \ref{null_dbf_dist} by conducting several multi-locus GWA studies on the Alzheimer Disease Neuroimaging Initiative (ADNI) cohort. The available dataset consists of $254$ subjects, $101$ cases of AD and $153$ controls, all genotyped at $316,348$ SNPs. We assume the genome has been partitioned into subsets of $P=5$ SNPs so that, for each subset, each of the $N$ individuals in the given study cohort is represented by the discrete-valued $P$-dimensional vector $\{\bm{y}_i=(y_{i1},\ldots,y_{iP})\}_{i=1}^N$. This results in a total number of $316,260$ SNP sets to be compared across the two populations. For each sliding window, five genetic distances (IBS, Simple Matching, Sokal and Sneath, Rogers and Tanimoto I and Hamman I) were applied, and the DBF statistic and corresponding p-value computed using the approach described in Section \ref{null_dbf_dist}. Due to the multiple testing problem, a genome-wide significance threshold of $10^{-7}$ was set to identify statistically significant SNP sets. 

 \begin{figure*}[h!]
 \center
 \includegraphics[scale=0.53]{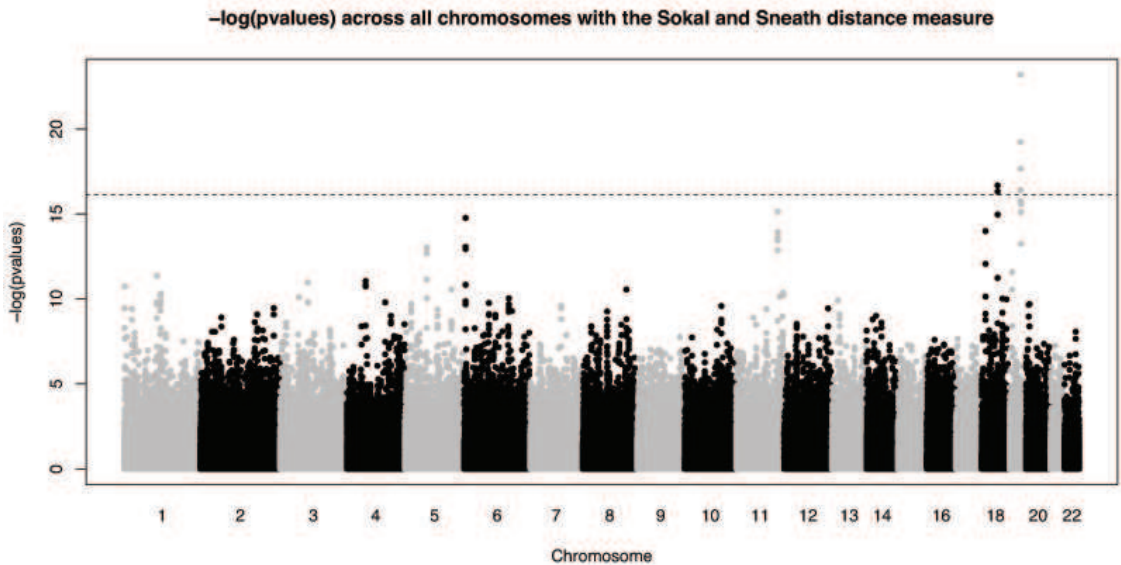}
 \caption{Manhattan plot of the -log(pvalues) computed across the genome for the Sokal and Sneath distance measure. Each point represents a window containing a multi-locus SNP subset consisting of $5$ adjacent SNPs. The dashed line represents the genome-wide significance threshold of $-\textrm{log}(10^{-7})$. The black and grey colours were used to distinguish between adjacent chromosomes.}
 \label{manhattan_ss}
 \end{figure*}

\begin{table*}[h!]
\begin{center} 
\caption{Significant SNPs and genes identified for each distance measure with the genome-wide significance threshold of $10^{-7}$. The chromosome in which the SNPs were identified are given, in addition to the p-value of the sliding window containing the given SNP.}
\label{tab:table_adni_results}
\begin{tabular}{c||c|l|c|c}
        \toprule
Distance measure & SNP&Gene&Chromosome &P-value of window\\
\midrule
IBS&$\begin{array}{l}\textrm{rs2075650}\\
\textrm{rs8106922}\\
\textrm{rs5167}\\
\textrm{apoe4}\\
\textrm{rs3760627}\\
\textrm{rs405509}\\
\textrm{rs2075642}\\
\textrm{rs6859}\\
\textrm{rs157580}\end{array}$&$\begin{array}{l}\textrm{TOMM40}\\
\textrm{TOMM40}\\
\textrm{APOC2, APOC4}\\
\textrm{APOE}\\
\textrm{CLPTM1}\\
\textrm{APOE}\\
\textrm{PVRL2}\\
\textrm{PVRL2}\\
\textrm{TOMM40}\end{array}$
&$\begin{array}{l}19\\
19\\
19\\
19\\
19\\
19\\
19\\
19\\
19\end{array}$
&$\begin{array}{l}1.626\times 10^{-10}\\
3.600\times 10^{-10}\\
4.844\times 10^{-10}\\
4.844\times 10^{-10}\\
1.237\times 10^{-9}\\
3.080\times 10^{-9}\\
6.449\times 10^{-8}\\
6.449\times 10^{-8}\\
6.449\times 10^{-8}\end{array}$\\
\midrule
Sokal and Sneath&$\begin{array}{l}\textrm{apoe4}  \\ 
\textrm{rs405509} \\
\textrm{rs2075650}\\ 
\textrm{rs8106922}\\  
\textrm{rs157580}\\   
\textrm{rs1222938}\\  
\textrm{rs12960771}\\ 
\textrm{rs1560531} \\ 
\textrm{rs2960617}\\  
\textrm{rs3862684} \\ 
\textrm{rs2075642} \\ 
\textrm{rs4803766}\\  
\textrm{rs6859}   \\  
\textrm{rs17748116}\end{array}$ &$\begin{array}{l}\textrm{APOE}  \\ 
\textrm{APOE} \\
\textrm{TOMM40}\\ 
\textrm{TOMM40}\\  
\textrm{TOMM40}\\   
\textrm{DCC}\\  
\textrm{DCC}\\ 
\textrm{DCC} \\ 
\textrm{DCC}\\  
\textrm{DCC} \\ 
\textrm{PVRL2} \\ 
\textrm{PVRL2}\\  
\textrm{PVRL2}   \\  
\textrm{DCC}\end{array}$ &$\begin{array}{l}19 \\ 
19\\
19 \\
19\\ 
19\\  
18\\   
18\\  
18\\ 
18 \\ 
18\\  
19 \\ 
19 \\ 
19\\  
18\end{array}$ &$\begin{array}{l}8.458\times 10^{-11}\\ 
8.458\times 10^{-11}\\
8.458\times 10^{-11}\\
8.458\times 10^{-11}\\ 
2.104\times 10^{-8}\\  
5.736\times 10^{-8}\\   
5.736\times 10^{-8}\\  
5.736\times 10^{-8}\\ 
5.736\times 10^{-8}\\ 
5.736\times 10^{-8}\\  
7.498\times 10^{-8}\\ 
7.498\times 10^{-8}\\ 
7.498\times 10^{-8}\\  
8.212\times 10^{-8}\end{array}$ 
\\
\midrule
$\begin{array}{l}\textrm{Rogers and} \\
\textrm{Tanimoto I}\end{array}$&$\begin{array}{l}\textrm{rs157580}  \\ 
\textrm{rs2075650} \\
\textrm{rs8106922}\\ 
\textrm{rs5167}\\  
\textrm{apoe4}\\   
\textrm{rs405509}\end{array}$ &$\begin{array}{l}\textrm{TOMM40}  \\ 
\textrm{TOMM40} \\
\textrm{TOMM40}\\ 
\textrm{APOC2, APOC4}\\  
\textrm{APOE}\\   
\textrm{APOE}\end{array}$ &$\begin{array}{l}19  \\ 
19 \\
19\\ 
19\\  
19\\   
19\end{array}$ &$\begin{array}{l}4.129\times 10^{-10} \\ 
4.129\times 10^{-10}\\
1.067\times 10^{-9}\\ 
6.915\times 10^{-8}\\  
6.915\times 10^{-8}\\   
6.915\times 10^{-8}\end{array}$\\
\midrule
$\begin{array}{l}\textrm{Simple Matching}\\\textrm{Hamman I}\end{array}$&$\begin{array}{l}\textrm{apoe4}\\
\textrm{rs405509}\\
\textrm{rs157580}\\
\textrm{rs2075650}\\
\textrm{rs8106922}\end{array}$
&$\begin{array}{l}\textrm{APOE}\\
\textrm{APOE}\\
\textrm{TOMM40}\\
\textrm{TOMM40}\\
\textrm{TOMM40}\end{array}$
&$\begin{array}{l}19\\
19\\
19\\
19\\
19\end{array}$&$\begin{array}{l}2.738\times 10^{-10}\\
2.738\times 10^{-10}\\
2.738\times 10^{-10}\\
2.738\times 10^{-10}\\
2.738\times 10^{-10}\end{array}$\\
\bottomrule
\end{tabular}
\end{center}
\end{table*}

In Figure \ref{manhattan_ss} we provide the corresponding Manhattan plot which depicts the significant SNP subsets across the entire genome for the Sokal and Sneath distance measure, showing the greatest effects around chromosomes $18$ and $19$. The results of all distance measures were summarized by the unique SNP and gene combinations identified; see Table \ref{tab:table_adni_results}. All significant SNPs are identified in chromosomes 18 and 19. In particular, chromosome 19 contains two genes, APOE4 and TOMM40, which are the major genetic variants found in many studies (see, for example, \cite{braskie2011neuroimaging} and \cite{shen2010whole}). Other previously reported genetic variants that overlap with our findings include APOC2, APOC4, PVRL2 and CLPTM1 \citep{takei2009genetic,yu2007comprehensive}. The DCC gene has also been previously identified \citep{bredesen2009neurodegeneration, lourenco2009netrin}.

\begin{figure*}[h!]
\center
\includegraphics[scale=0.745]{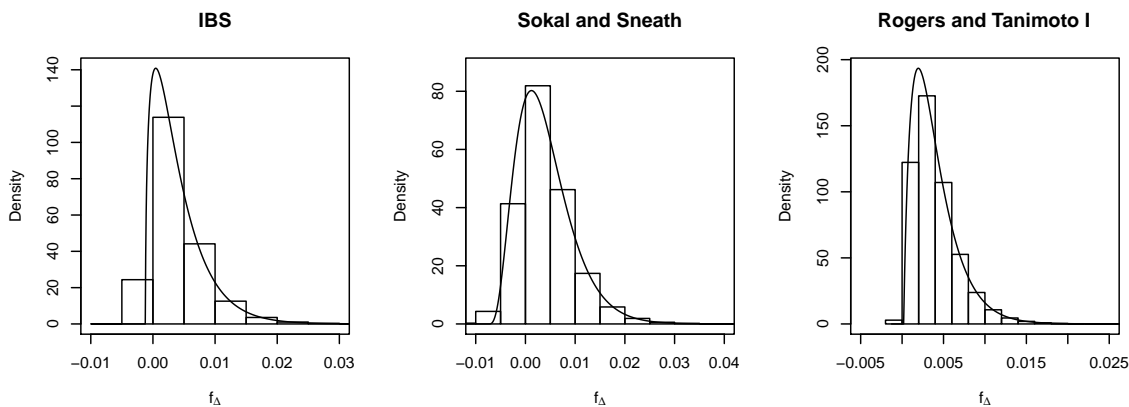}
\caption{The approximate PDF of $F_{\bm{\Delta}}$ for the IBS, Sokal and Sneath and Rogers and Tanimoto I distance measures versus the sampling values of $F_{\bm{\Delta}}$ obtained with $10^6$ Monte Carlo permutations for the first sliding window of SNPs for chromosome $19$. This window contains $254$ samples of the $5$ SNPs rs12459906, rs11878315, rs10409452, rs3813154 and rs895344.}
\label{5_dists_pdfs_hists}
\end{figure*}

As an illustrative comparison with the permutation approach, in Figure \ref{5_dists_pdfs_hists} we compare the approximate distribution of the DBF statistic under the null with the null permutation distribution obtained using by $10^6$ Monte Carlo permutations. This was done for the first sliding window of chromosome 19 and three genetic distances. Here again it can be clearly seen that the approximate null distribution of the DBF statistic provides a good fit. 

\begin{figure*}[h!]
\center
\includegraphics[scale=0.55]{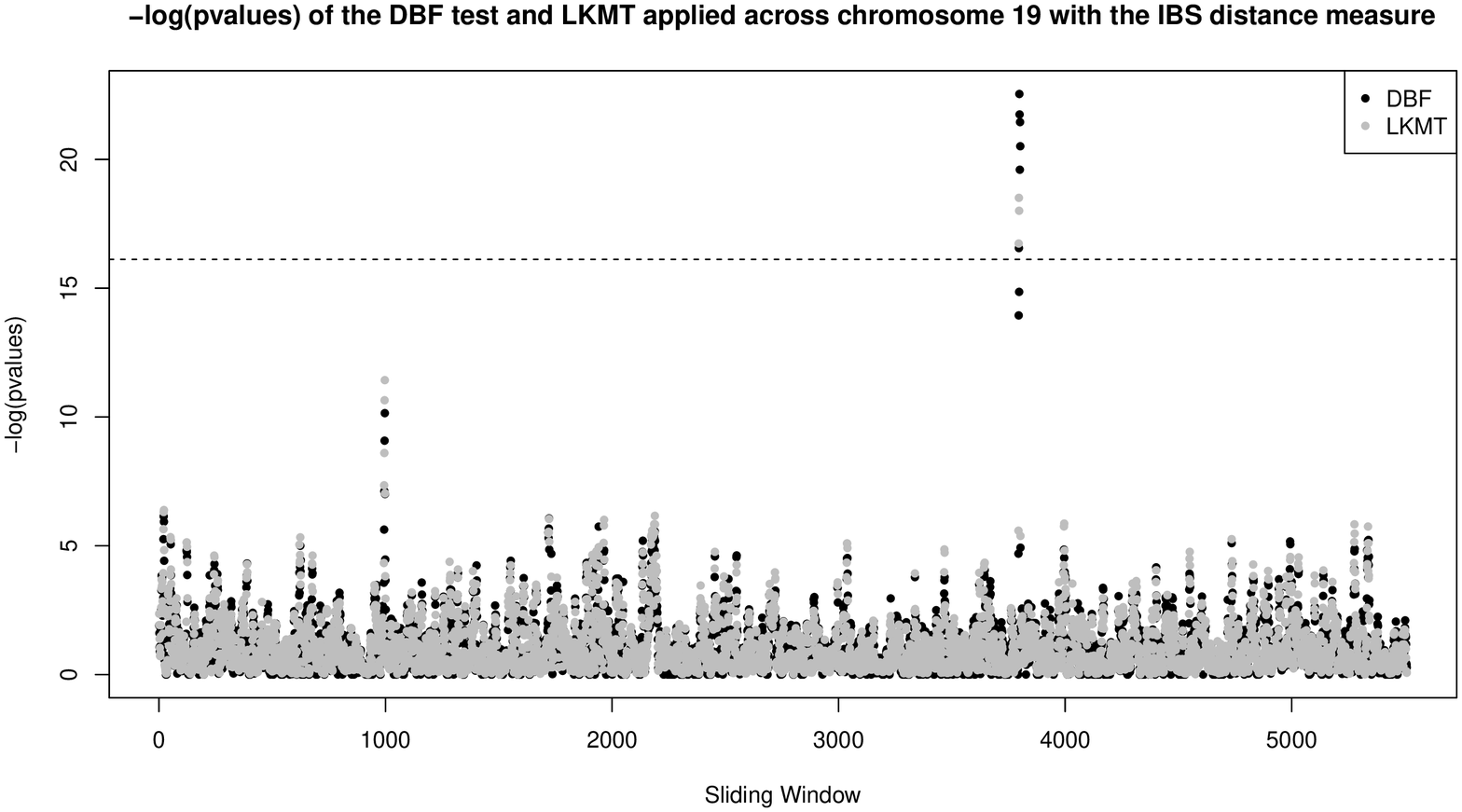}
\caption{ $-$log(pvalues) computed across chromosome 19 using the DBF and LKMT tests with the IBS distance measure and kernel function, respectively. Each point represents a multilocus SNP subset consisting of $5$ adjacent SNPs, and the dashed line represents the transformed genome-wide significance threshold of $-\textrm{log}(10^{-7})$.}
\label{compar_dbf_lkmt}
\end{figure*}

We have also compared our results with those obtained using a statistical procedure developed specifically for multi-locus case-control studies, the kernel-machine logistic regression approach of \cite{wu2010powerful}, which we denote LKMT. This method makes use of similarities rather than distances via kernel functions $K(\cdot,\cdot)$; note that kernels are related to distances, for example, the IBS kernel between individuals $i$ and $j$ is just one minus their IBS distance. Since LKMT makes use of an approximate distribution rather than permutations, we are able to compare the two methods without major computational issues. LKMT consists of a logistic kernel-machine regression model where the probability that a given subject is a case subject is modelled as some function of the similarities between that subject and all other subjects. We denote the case-control status of the $N$ subjects by $\{s_i\}_{i=1}^N$, where $s_i=0$ for controls and $s_i=1$ for cases. The model is then given by $\textrm{logit}[P(s_i=1)]=\beta_0+m(\bm{y}_i)$ for $i=1,\ldots,N$, where $\beta_0$ is an intercept, and $m(\bm{z}_i)=\sum_{j=1}^N \delta_j K(\bm{y}_i,\bm{y}_j)$ for some constants $\{\delta_j\}_{j=1}^N$ so that $m(\cdot)$ is completely specified by the similarities between the subjects depicted by the kernel function $K(\cdot,\cdot)$. The null hypothesis of no group effect is stated as $H_0: h(\bm{y}_i)=0$ for $i=1,\ldots,N$, and the variance-component score statistic $Q=(\bm{s}-\hat{\bm{\beta}}_0)^T\bm{K}(\bm{s}-\hat{\bm{\beta}}_0)/2$
is used to test this, where $\bm{s}=(s_1,\ldots,s_N)^T$, $\hat{\bm{\beta}_0}=\hat{\beta}_0(1,\ldots,1)^T$ and $\bm{K}=\{K(\bm{y}_i,\bm{y}_j)\}_{i,j=1}^N$. The distribution of $Q$ under the null hypothesis is approximated by a Chi-squared distribution.

The p-values obtained by LKMT are of the same order as our approach with the IBS distance measure when using the comparable IBS kernel. For instance, for chromosome 19, which is the chromosome containing the smallest p-values, Figure \ref{compar_dbf_lkmt} provides a visual comparison of the two methods when using the IBS measure; similar results are obtained for other chromosomes  (not shown). Although a rigorous power comparison is beyond the scope of this study, it can be noted that both methods identified the same causal SNPs in Chromosome 19 at the significance level of $10^{-7}$, i.e., the ones listed in Table \ref{tab:table_adni_results}. Thus the well-known APOE and TOMM40 genes were identified by both approaches.

\section{Conclusion} \label{conclusion}

For large datasets, such as those arising in biological applications, it is desirable to apply distance-based tests for equality between groups without permutations. This requires that the discrete sampling distribution of the chosen statistic under the null is approximated by a suitably chosen continuous distribution. In this article we focused on the DBF statistic due to the intuitive distance-based variance decomposition it uses. We described some results relating the DBF statistic to classical MANOVA and ANOVA statistics. Its connection with ANOVA is well-known, allowing an exact null distribution of the DBF statistic to be obtained when applied in the special case of univariate Normal data with the Euclidean distance measure. Its connection with MANOVA, however, is not so well-known. We presented the result that for multivariate Normal data with $G=2$, the DBF statistic applied with a Mahalanobis-like distance measure is related to Hotelling's $T^2$ statistic. From this connection the exact null distribution of the transformed DBF statistic is attainable.

For more general data types, and corresponding distance measures, where the exact null distribution of the DBF statistic is unknown, we considered its permutation distribution. We showed that it depends on the permutation distribution of the between-group variability component of the distance-based variance decomposition. On presenting the skewed characteristics of the between-group variability for real biological datasets, we justified the use of the Pearson type III distribution to model its skewed nature. We then used its monotonic relationship with the DBF statistic to derive an approximate distribution for the DBF statistic. 

For a range of data types and distance measures we provided evidence to support the performance of the proposed approximation. We also used it to conduct several case-control GWA studies on the ADNI dataset in order to show its applicability to a large dataset. This study identified a number of genes inline with validated genetic risk factors for Alzheimer's disease. We also showed that the proposed approach performs comparably to an existing test statistic that has been specifically designed for case-control association studies.  

The R software for performing the DBF test with the approximate null distribution is available from {\tt http://www2.imperial.ac.uk/$\sim$gmontana}.


\section*{Acknowledgements}

Data collection and sharing for this project was funded by the Alzheimer's Disease Neuroimaging Initiative (ADNI) (National Institutes of Health Grant U01 AG024904). ADNI is funded by the National Institute on Aging, the National Institute of Biomedical Imaging and Bioengineering, and through generous contributions from the following: Abbott; Alzheimer’s Association; Alzheimer’s Drug Discovery Foundation; Amorfix Life Sciences Ltd.; AstraZeneca; Bayer HealthCare; BioClinica,Inc.; Biogen Idec Inc.;Bristol-Myers Squibb Company; Eisai Inc.; Elan Pharmaceuticals Inc.; Eli Lilly and Company; F. Hoffmann-La Roche Ltd and its affiliated company Genentech, Inc.;GE Healthcare;Innogenetics, N.V.;Janssen Alzheimer Immunotherapy Research $\&$ Development, LLC.;Johnson $\&$ Johnson Pharmaceutical Research $\&$ Development LLC.; Medpace, Inc.; Merck $\&$ Co., Inc.; Meso Scale Diagnostics, LLC.; Novartis Pharmaceuticals Corporation; Pfizer Inc.; Servier; Synarc Inc.; and Takeda Pharmaceutical Company. The Canadian Institutes of Health Research is providing funds to support ADNI clinical sites in Canada. Private sector contributions are facilitated by the Foundation for the National Institutes of Health (www.fnih.org). The grantee organization is the Northern California Institute for Research and Education, and the study is coordinated by the Alzheimer's Disease Cooperative Study at the University of California, San Diego. ADNI data are disseminated by the Laboratory for Neuro Imaging at the University of California, Los Angeles. This research was also supported by NIH grants P30 AG010129 and K01 AG030514.

\section*{Appendix}
\begin{appendix}
                
\section{Multi-locus genetic distance measures}
Assume $N$ individuals in the given study cohort are observed at $P$ SNPs, yielding the discrete-valued $P$-dimensional vectors $\{\bm{y}_i=(y_{i1},\ldots,y_{iP})\}_{i=1}^N$. The identity-by-state (IBS) distance measure is commonly used, and measures the proportion of risk alleles (minor alleles) shared between individuals across the SNPs considered \citep{wessel2006generalized,wu2010powerful,mukhopadhyay2010association}. The distance between individuals $i$ and $j$ is defined by
\begin{equation}\nonumber
d_{IBS}(\bm{y}_i,\bm{y}_j)=1-\frac{1}{2P}\sum_{k=1}^Ps(y_{ik},y_{jk}),
\end{equation}
where $s(y_{ik},y_{jk})=0$ if $y_{ik}=0$ and $y_{jk}=2$, or if $y_{ik}=2$ and $y_{jk}=0$, $s(y_{ik},y_{jk})=1$ if $y_{ik}=1$ and $y_{jk}\neq 1$, or if $y_{jk}=1$ and $y_{ik}\neq 1$, and $s(y_{ik},y_{jk})=2$ if $y_{ik}=y_{jk}$. This distance takes values between $0$ and $1$. Weighted versions of this distance also exist where a weight is attached to each of the $P$ SNPs depending on properties such as functional significance or frequency of the minor allele \citep{wessel2006generalized,li2009genetic}. Genetic distances have also been proposed based on the contingency table between individuals $i$ and $j$ containing the frequency that each combination of $y_{ik}$ and $y_{jk}$ values occur over the range of SNPs considered \citep{selinski2005cluster}; see Table \ref{tab_snp_dist} below. 
 
\begin{table*}[h!]
\begin{center} 
\caption{Contingency table containing the frequency of a given combination of minor allele count between individuals $i$ and $j$ over the $P$ SNPs. $m_{kl}$ is the frequency of individual $i$ having $k$ minor alleles and individual $j$ having $l$ minor alleles.}
\label{tab_snp_dist}
\begin{tabular}{c|ccccc}
        \toprule
& \multicolumn{1}{c}{}& \multicolumn{1}{c}{Individual $j$}&  \multicolumn{1}{c}{}\\
Individual $i$ & $0$&$1$&$2$  \\
\midrule
 $0$&$m_{00}$&$m_{01}$&$m_{02}$ \\      
 $1$&$m_{10}$&$m_{11}$&$m_{12}$ \\      
 $2$&$m_{20}$&$m_{21}$&$m_{22}$ \\ 
\bottomrule
\end{tabular}
\end{center}

\end{table*}

The key statistics in this table are the number of complete matches of the minor alleles, $m^{+}=\sum_{k=0}^2m_{kk}$, and the number of mismatches, $m^{-}=P - m^{+}$, where the total number of comparisons made is $P$. Based on these quantities, the following `matching coefficient' distance measures can be defined; the Simple Matching distance 
\begin{equation}
d_{SM}(\bm{y}_i,\bm{y}_j)=1-\frac{m^{+}}{P},\nonumber
\end{equation}
the Sokal and Sneath distance
\begin{equation}
d_{SS}(\bm{y}_i,\bm{y}_j)=1-\frac{m^{+}}{m^{+}+\frac{1}{2}m^{-}},\nonumber
\end{equation}
the Rogers and Tanimoto I distance 
\begin{equation}
d_{RTI}(\bm{y}_i,\bm{y}_j)=1-\frac{m^{+}}{m^{+}+2m^{-}},\nonumber
\end{equation}
and the Hamman I distance 
\begin{equation}
d_{HI}(\bm{y}_i,\bm{y}_j)=1-\frac{s^*(\bm{y}_i,\bm{y}_j)}{\max_{i,j}\{s^*(\bm{y}_i,\bm{y}_j)\}},\nonumber
\end{equation}
where $s^*(\bm{y}_i,\bm{y}_j)=s_{HI}(\bm{y}_i,\bm{y}_j)+|\min_{i,j}\{s_{HI}(\bm{y}_i,\bm{y}_j)\}|$ with $s_{HI}(\bm{y}_i,\bm{y}_j)=(m^{+}-m^{-})/P$. All of these distances take values between $0$ and $1$.
        
\section{Distance measures between curves}

Assume a set of $N$ curves $\{y_i(t)\}_{i=1}^N$ defined for $t\in\tau$ where all curves have been defined over the same range $\tau$. The L$_2$ distance represents the area between curves, and hence the magnitude of difference between them \citep{ferraty2006nonparametric,minas2011distance,salem2010curve}, and is defined by 
\begin{equation}
d_L(y_i,y_j) = \left(\int_{\tau}\left(y_i(t)-y_j(t)\right)^2 dt\right)^{\frac{1}{2}}.\nonumber
\end{equation}

The curvature distance  quantifies the difference in the rate of change between two curves \citep{ferraty2006nonparametric,minas2011distance}, and is defined by
\begin{equation}
d_C(y_i,y_j) =\left|\int_{\tau}\left(y_i^{\prime\prime}(t)\right)^2dt-\int_{\tau}\left(y_j^{\prime\prime}(t)\right)^2dt\right|.\nonumber
\end{equation}

The visual L$_2$ distance  quantifies the difference in the scale-invariant shape between curves, analogously to the difference detected by the human eye \citep{Marron1995,minas2011distance}. For this distance, curves $y_i$ and $y_j$ are scaled both in time and magnitude, so that their values range between $0$ and $1$ in time period $[0,1]$; denote these $y^s_i(t)$ and $y^s_j(t)$ where $t\in[0,1]$. They are then represented as infinite sets of points in the two-dimensional plane, denoted $p_i = \left\{(t,y^s_i(t)) ~ | ~ t\in[0,1]\right\}$ and $p_j = \left\{(t,y^s_j(t)) ~ | ~ t\in[0,1]\right\}$. The visual L$_2$ distance is then defined by
\begin{equation}
d_V(y_i,y_j)=\left(\int_0^1 \delta_{ij}^2(t) dt +\int_0^1 \delta_{ji}^2(t) dt\right)^{\frac{1}{2}},\nonumber
\end{equation}
where $\delta_{ij}(t)$ is the minimum Euclidean distance between the point $y^s_i(t)$ and all points $p_j$ representing $y^s_j$; note that $\delta_{ij}(t)$ is not necessarily equal to $\delta_{ji}(t)$.
        
\section{Derivation of the CDF of $F_{\Delta}$ for $\gamma_B>0$}

First consider the case where $-\infty<f<-1$. By inspection of Figure \ref{all_skew_alpha} (a), we see that we need only consider the relationship between $F_{\bm{\Delta}}$ and $B^s_{\bm{\Delta}}$ where $B^s_{\bm{\Delta}} > \beta $. We thus have that 
\begin{eqnarray}
\mathcal{F}_{F_{\bm{\Delta}}}\left(f;\mu_B,\sigma_B,\gamma_B\right)&=&P(F_{\bm{\Delta}}\leq f;\mu_B,\sigma_B,\gamma_B)\nonumber\\
&=&P\left(\beta< B^s_{\bm{\Delta}}\leq h^{-1}(f);\gamma_B\right)\nonumber\\
&=&P\left(B^s_{\bm{\Delta}}\leq h^{-1}(f);\gamma_B\right) - \left(B^s_{\bm{\Delta}}\leq \beta;\gamma_B\right)\nonumber\\
&=&\mathcal{F}_{B^s_{\bm{\Delta}}}\left( h^{-1}(f);\gamma_B\right) - \mathcal{F}_{B^s_{\bm{\Delta}}}\left(\beta;\gamma_B\right).\nonumber
\end{eqnarray}

Now consider the case where $\alpha\leq f < \infty$. From Figure \ref{all_skew_alpha} (a) we see that we must consider the relationship between $F_{\bm{\Delta}}$ and $B^s_{\bm{\Delta}}$ for $B^s_{\bm{\Delta}}<\beta$, while adding the cumulative component of all values of $F_{\bm{\Delta}}$ for which $B^s_{\bm{\Delta}}>\beta$. That is, 
\begin{eqnarray}
\mathcal{F}_{F_{\bm{\Delta}}}\left(f;\mu_B,\sigma_B,\gamma_B\right)&=&P\left(\alpha\leq F_{\bm{\Delta}}\leq f;\mu_B,\sigma_B,\gamma_B\right)+P(-\infty<F_{\bm{\Delta}}< -1;\mu_B,\sigma_B,\gamma_B)\nonumber\\
&=&P\left(\frac{-2}{\gamma_B}\leq B^s_{\bm{\Delta}}\leq h^{-1}(f);\gamma_B\right)+P\left(\beta< B^s_{\bm{\Delta}}< h^{-1}(-1);\gamma_B\right)\nonumber\\
&=&\mathcal{F}_{B^s_{\bm{\Delta}}}\left( h^{-1}(f);\gamma_B\right) - \mathcal{F}_{B^s_{\bm{\Delta}}}\left(\frac{-2}{\gamma_B};\gamma_B\right)+P\left(\beta< B^s_{\bm{\Delta}}< \infty;\gamma_B\right)\nonumber\\
&=&\mathcal{F}_{B^s_{\bm{\Delta}}}\left( h^{-1}(f);\gamma_B\right) - \mathcal{F}_{B^s_{\bm{\Delta}}}\left(\frac{-2}{\gamma_B};\gamma_B\right)+\mathcal{F}_{B^s_{\bm{\Delta}}}\left(\infty;\gamma_B\right)-\mathcal{F}_{B^s_{\bm{\Delta}}}\left(\beta;\gamma_B\right),\nonumber
\end{eqnarray}
and since $-2/\gamma_T\leq B^s_{\bm{\Delta}}<\infty$, we have $\mathcal{F}_{B^s_{\bm{\Delta}}}\left(-2/\gamma_B;\gamma_B\right)=0$ and $\mathcal{F}_{B^s_{\bm{\Delta}}}\left(\infty;\gamma_B\right)=1$, so that 
\begin{equation}
\mathcal{F}_{F_{\bm{\Delta}}}\left(f;\mu_B,\sigma_B,\gamma_B\right)=1+\mathcal{F}_{B^s_{\bm{\Delta}}}\left(h^{-1}(f);\gamma_B\right)-\mathcal{F}_{B^s_{\bm{\Delta}}}\left(\beta;\gamma_B\right).\nonumber
\end{equation}
Thus we have that the cumulative distribution function of $F_{\bm{\Delta}}$ is given by
\begin{equation}\nonumber
\mathcal{F}_{F_{\bm{\Delta}}}\left(f;\mu_B,\sigma_B,\gamma_B\right)=\begin{cases}
\mathcal{F}_{B^s_{\bm{\Delta}}}\left( h^{-1}(f);\gamma_B\right)-\mathcal{F}_{B^s_{\bm{\Delta}}}\left(\beta;\gamma_B\right)& -\infty<f<-1 \\
1+\mathcal{F}_{B^s_{\bm{\Delta}}}\left( h^{-1}(f);\gamma_B\right)-\mathcal{F}_{B^s_{\bm{\Delta}}}\left(\beta;\gamma_B\right)& \alpha\leq f<\infty
\end{cases}
\end{equation}
for $\gamma_B>0$, as required.

Next we show that this is a valid CDF by showing that the following conditions are satisfied:
\begin{enumerate}[(i)]
\item The limit of $\mathcal{F}_{F_{\bm{\Delta}}}(f)$ as $f$ tends to $-\infty$ from the right is $0$, and as $f$ tends to $\infty$ from the left is $1$. That is
\begin{equation}
\lim_{f\to -\infty^+}{\left[\mathcal{F}_{F_{\bm{\Delta}}}\left(f;\mu_B,\sigma_B,\gamma_B\right)\right]}=0\quad\textrm{and}\quad \lim_{f\to \infty^-}{\left[\mathcal{F}_{F_{\bm{\Delta}}}\left(f;\mu_B,\sigma_B,\gamma_B\right)\right]}=1.\nonumber
\end{equation}

These follow because 
\begin{eqnarray}
\lim_{f\to -\infty^+}{\left[\mathcal{F}_{F_{\bm{\Delta}}}\left(f;\mu_B,\sigma_B,\gamma_B\right)\right]}&=&\lim_{b\to \beta^+}\left[\mathcal{F}_{B^s_{\bm{\Delta}}}(b;\gamma_B)\right]-\mathcal{F}_{B^s_{\bm{\Delta}}}(\beta;\gamma_B)\nonumber\\
&=&\mathcal{F}_{B^s_{\bm{\Delta}}}(\beta;\gamma_B)-\mathcal{F}_{B^s_{\bm{\Delta}}}(\beta;\gamma_B)\nonumber\\
&=&0,\nonumber
\end{eqnarray}
and
\begin{eqnarray}
\lim_{f\to \infty^-}{\left[\mathcal{F}_{F_{\bm{\Delta}}}\left(f;\mu_B,\sigma_B,\gamma_B\right)\right]}&=&1+\lim_{b\to \beta^-}\left[\mathcal{F}_{B^s_{\bm{\Delta}}}(b;\gamma_B)\right]-\mathcal{F}_{B^s_{\bm{\Delta}}}(\beta;\gamma_B)\nonumber\\
&=&1+\mathcal{F}_{B^s_{\bm{\Delta}}}(\beta;\gamma_B)-\mathcal{F}_{B^s_{\bm{\Delta}}}(\beta;\gamma_B)\nonumber\\
&=&1.\nonumber
\end{eqnarray}

\item $\mathcal{F}_{F_{\bm{\Delta}}}(f)$ is a monotone, non-decreasing function of $f$. That is, for $f_1< f_2$, $\mathcal{F}_{F_{\bm{\Delta}}}(f_1) \leq \mathcal{F}_{F_{\bm{\Delta}}}(f_2)$.

For $-\infty<f_1<f_2<-1$, we have that
\begin{eqnarray}
\mathcal{F}_{F_{\bm{\Delta}}}\left(f_1;\mu_B,\sigma_B,\gamma_B\right)&=&\mathcal{F}_{B^s_{\bm{\Delta}}}\left( h^{-1}(f_1);\gamma_B\right)-\mathcal{F}_{B^s_{\bm{\Delta}}}\left(\beta;\gamma_B\right)\nonumber\\
\mathcal{F}_{F_{\bm{\Delta}}}\left(f_2;{\mu}_B,\sigma_B,\gamma_B\right)&=&\mathcal{F}_{B^s_{\bm{\Delta}}}\left( h^{-1}(f_2);\gamma_B\right)-\mathcal{F}_{B^s_{\bm{\Delta}}}\left(\beta;\gamma_B\right),\nonumber
\end{eqnarray}
so that $\mathcal{F}_{F_{\bm{\Delta}}}\left(f_1;\mu_B,\sigma_B,\gamma_B\right)-\mathcal{F}_{F_{\bm{\Delta}}}\left(f_2;\mu_B,\sigma_B,\gamma_B\right)$ is equal to
\begin{equation}
\mathcal{F}_{B^s_{\bm{\Delta}}}\left( h^{-1}(f_1);\gamma_B\right)-\mathcal{F}_{B^s_{\bm{\Delta}}}\left( h^{-1}(f_2);\gamma_B\right).\nonumber
\end{equation}
This is negative since $h^{-1}(f_1)<h^{-1}(f_2)$ and $\mathcal{F}_{B^s_{\bm{\Delta}}}(b;\gamma_B)$ is a non-decreasing, monotone function of $b$ (as it is a valid CDF). Hence $\mathcal{F}_{F_{\bm{\Delta}}}(f_1) \leq \mathcal{F}_{F_{\bm{\Delta}}}(f_2)$, as required.

For $\alpha\leq f_1<f_2<\infty$, we have that 
\begin{eqnarray}
\mathcal{F}_{F_{\bm{\Delta}}}\left(f_1;\mu_B,\sigma_B,\gamma_B\right)&=&1+\mathcal{F}_{B^s_{\bm{\Delta}}}\left( h^{-1}(f_1);\gamma_B\right)-\mathcal{F}_{B^s_{\bm{\Delta}}}\left(\beta;\gamma_B\right)\nonumber\\
\mathcal{F}_{F_{\bm{\Delta}}}\left(f_2;\mu_B,\sigma_B,\gamma_B\right)&=&1+\mathcal{F}_{B^s_{\bm{\Delta}}}\left( h^{-1}(f_2);\gamma_B\right)-\mathcal{F}_{B^s_{\bm{\Delta}}}\left(\beta;\gamma_B\right),\nonumber
\end{eqnarray}
so that $\mathcal{F}_{F_{\bm{\Delta}}}\left(f_1;\mu_B,\sigma_B,\gamma_B\right)-\mathcal{F}_{F_{\bm{\Delta}}}\left(f_2;\mu_B,\sigma_B,\gamma_B\right)$ is equal to
\begin{equation}
\mathcal{F}_{B^s_{\bm{\Delta}}}\left( h^{-1}(f_1);\gamma_B\right)-\mathcal{F}_{B^s_{\bm{\Delta}}}\left( h^{-1}(f_2);\gamma_B\right).\nonumber
\end{equation}
As before, this is negative since $h^{-1}(f_1)<h^{-1}(f_2)$ and $\mathcal{F}_{B^s_{\bm{\Delta}}}(b;\gamma_B)$ is non-decreasing and monotone.  Hence $\mathcal{F}_{F_{\bm{\Delta}}}(f_1) \leq \mathcal{F}_{F_{\bm{\Delta}}}(f_2)$, as required.

Finally, let $f_1=-1$ and $f_2=\alpha$. Then 
\begin{eqnarray}
\mathcal{F}_{F_{\bm{\Delta}}}\left(f_1;\mu_B,\sigma_B,\gamma_B\right)&=&1-\mathcal{F}_{B^s_{\bm{\Delta}}}\left(\beta;\gamma_B\right)\nonumber\\
\mathcal{F}_{F_{\bm{\Delta}}}\left(f_2;\mu_B,\sigma_B,\gamma_B\right)&=&1+\mathcal{F}_{B^s_{\bm{\Delta}}}\left( \frac{-2}{\gamma_B};\gamma_B\right)-\mathcal{F}_{B^s_{\bm{\Delta}}}\left(\beta;\gamma_B\right),\nonumber
\end{eqnarray}
and since $\mathcal{F}_{B^s_{\bm{\Delta}}}\left(-2/\gamma_B;\gamma_B\right)=0$, we have that $\mathcal{F}_{F_{\bm{\Delta}}}(f_1) \leq \mathcal{F}_{F_{\bm{\Delta}}}(f_2)$ holds at the discontinuity.

\item $\mathcal{F}_{F_{\bm{\Delta}}}(f)$ is continuous from the right, that is
\begin{equation}
\lim_{\epsilon\to 0^+}{\left[\mathcal{F}_{F_{\bm{\Delta}}}\left(f+\epsilon;\mu_B,\sigma_B,\gamma_B\right)\right]}=\mathcal{F}_{F_{\bm{\Delta}}}\left(f;\mu_B,\sigma_B,\gamma_B\right).\nonumber
\end{equation}

For $-\infty<f<-1$ we have that 
\begin{eqnarray}
\lim_{\epsilon\to 0^+}{\left[\mathcal{F}_{F_{\bm{\Delta}}}\left(f+\epsilon;\mu_B,\sigma_B,\gamma_B\right)\right]}&=&
\lim_{\epsilon\to 0^+}{\left[\mathcal{F}_{B^s_{\bm{\Delta}}}\left( h^{-1}(f+\epsilon);\gamma_B\right)\right]}-\mathcal{F}_{B^s_{\bm{\Delta}}}\left(\beta;\gamma_B\right)\nonumber\\
&=&\mathcal{F}_{B^s_{\bm{\Delta}}}\left( h^{-1}(f);\gamma_B\right)-\mathcal{F}_{B^s_{\bm{\Delta}}}\left(\beta;\gamma_B\right)\nonumber\\
&=&\mathcal{F}_{F_{\bm{\Delta}}}\left(f;\mu_B,\sigma_B,\gamma_B\right),\nonumber
\end{eqnarray}
and for $\alpha\leq f<\infty$ we have that 
\begin{eqnarray}
\lim_{\epsilon\to 0^+}{\left[\mathcal{F}_{F_{\bm{\Delta}}}\left(f+\epsilon;{\mu}_B,{\sigma}_B,{\gamma}_B\right)\right]}&=&
1+\lim_{\epsilon\to 0^+}{\left[\mathcal{F}_{B^s_{\bm{\Delta}}}\left( h^{-1}(f+\epsilon);{\gamma}_B\right)\right]}-\mathcal{F}_{B^s_{\bm{\Delta}}}\left(\beta;{\gamma}_B\right)\nonumber\\
&=&1+\mathcal{F}_{B^s_{\bm{\Delta}}}\left( h^{-1}(f);{\gamma}_B\right)-\mathcal{F}_{B^s_{\bm{\Delta}}}\left(\beta;{\gamma}_B\right)\nonumber\\
&=&\mathcal{F}_{F_{\bm{\Delta}}}\left(f;{\mu}_B,{\sigma}_B,{\gamma}_B\right),\nonumber
\end{eqnarray}
as required.
\end{enumerate}
Thus $\mathcal{F}_{F_{\bm{\Delta}}}\left(f;{\mu}_B,{\sigma}_B,{\gamma}_B\right)$ is a valid CDF for $\gamma_B>0$.
   
\end{appendix}  

\bibliography{bib_pseudo_f_dist_paper-1}
\bibliographystyle{natbib}
\end{document}